\documentclass[onecolumn,showpacs,preprintnumbers,amsmath,amssymb,aps,pre]{revtex4}

\usepackage{graphicx,color}

\begin{document}

\preprint{}

\title{Statistical mechanics of Beltrami flows in axisymmetric geometry:\\ Equilibria and bifurcations}

\author{Aurore Naso$^{1,2}$}
\author{Simon Thalabard$^{1,2}$}
\author{Gilles Collette$^{1,3}$}
\author{Pierre-Henri Chavanis$^{4}$}
\author{B\'ereng\`ere Dubrulle$^{1}$}

\affiliation{$^1$ SPEC/IRAMIS/CEA Saclay, and CNRS (URA 2464), 91191 Gif-sur-Yvette Cedex, France\\
$^2$ Laboratoire de Physique, Ecole Normale Sup\'erieure de Lyon and CNRS (UMR 5672),\\ 46, all\'ee d'Italie, 69364 Lyon Cedex 07, France\\
$^3$ IPSL/Laboratoire des Sciences du Climat et de l'Environnement, CEA-CNRS, 91191 Gif sur Yvette, France\\
$^4$ Laboratoire de Physique Th\'eorique (UMR 5152), Universit\'e Paul Sabatier, 118 route de Narbonne, 31062 Toulouse, France}

\begin{abstract}
We characterize the thermodynamical equilibrium
states of axisymmetric Euler-Beltrami flows. They have the form of coherent structures presenting
one or several cells. We find
the relevant control parameters and derive the corresponding equations of state. We
prove the coexistence of several equilibrium states
 for a given value of the control parameter like
in 2D turbulence [Chavanis \& Sommeria, J. Fluid Mech. {\bf 314}, 267
(1996)]. We explore the stability of these equilibrium states and
show that all states are saddle points
of entropy and can, in principle, be destabilized by
a perturbation with a larger wavenumber, resulting in a structure at
the smallest available scale. This mechanism is
therefore reminiscent of the 3D Richardson energy cascade
towards smaller and smaller scales. Therefore, our system is truly intermediate between 2D turbulence (coherent structures) and 3D turbulence (energy cascade). We further
explore numerically the robustness of the equilibrium
states with respect to random perturbations using a relaxation
algorithm in both canonical and microcanonical
ensembles. We show that saddle points of entropy can
be very robust and therefore play a role in the dynamics.
We evidence differences in the robustness of the
solutions in the canonical and microcanonical ensembles.  A scenario
of bifurcation between two different equilibria (with
one or two cells) is proposed and discussed in connection with a
recent observation of a turbulent bifurcation in a von K\'arm\'an
experiment [Ravelet {\it et al.}, Phys. Rev. Lett. {\bf 93}, 164501
(2004)].
\end{abstract}

\date{\today}

\maketitle

\section{Introduction}

The statistical mechanics of systems with long-range interactions has
recently attracted a lot of attention \cite{dauxois02}.  Typical systems with
long-range interactions include self-gravitating systems \cite{paddy,ijmpb}, two-dimensional
vortices \cite{chavanis02}, non-neutral plasmas \cite{dubinoneil}, free
electrons lasers \cite{barre04} and toy models such as the Hamiltonian
Mean Field (HMF) model \cite{ar,cvb}. Unusual
properties of systems with long-range interaction
such as negative specific heats or ensembles inequivalence
have been evidenced and linked with lack of additivity \cite{cdr}. In addition, a
striking property of these systems is the rapid formation of quasi stationary
self-organized states (coherent structures) such as galaxies in the universe \cite{bt}, large scale vortices  in geophysical and astrophysical flows \cite{flierl,marcus,williams,tabeling,cvh} or quasi-stationary states in the HMF model \cite{lrt,yamaguchi,antoniazzi}. These QSSs can be explained in terms of statistical mechanics using the theory developed by Lynden-Bell \cite{lb} for the Vlasov equation or by Miller \cite{miller} and Robert \& Sommeria \cite{rs} for the 2D Euler equation.

Two-dimensional vortices interact via a logarithmic potential. Interaction of vortices in 3D turbulence
is weaker than in 2D turbulence, but still long-range. Due to
dissipative anomaly and vortex stretching, statistical mechanics of 3D
turbulence has so far eluded theories. Recent progress was
recently made considering 3D inviscid axisymmetric flows
\cite{leprovost06,naso09} that are intermediate between 2D and 3D
flows: they are subject to vortex stretching like in 3D turbulence,
but locally conserve a scalar quantity in the ideal limit, like in 2D
turbulence. It is therefore interesting to study whether these systems
obey the peculiarities observed in other systems with long-range interactions such
 as violent relaxation, existence of long-lived quasi-stationary
states, negative specific heats and ensembles inequivalence.

The general study of the stability of axisymmetric
flows, and the possible occurrence of phase transitions,
is  difficult due to the presence of an infinite number
of Casimir invariants linked with the axisymmetry of the flow.  In a previous
paper \cite{naso09}, hereafter Paper I, we have considered a simplified
axisymmetric Euler system characterized by only three conserved
quantities: the fine-grained energy $E^{f.g.}$, the helicity
$H$ and the angular momentum $I$. We have developed the corresponding
statistical mechanics and shown that equilibrium states of this system
have the form of Beltrami mean flows on which are superimposed  Gaussian
fluctuations. We have  shown that the maximization of entropy $S$
at fixed helicity $H$, angular momentum $I$ and microscopic energy
$E^{f.g.}$ (microcanonical ensemble) is equivalent to the maximization of free
energy $J=S-\beta E^{f.g.}$ at fixed helicity $H$ and angular momentum
$I$ (canonical ensemble). These variational principles are also equivalent to the minimization of
macroscopic energy $E^{c.g.}$ at fixed helicity $H$ and angular
momentum $I$. This provides a justification of the minimum energy
principle  (selective decay) from statistical mechanics. We have furthermore discussed the analogy with the
simplified thermodynamical approach of 2D turbulence developed in \cite{naso09a}
based on only three conserved
quantities: the fine-grained enstrophy $\Gamma_2^{f.g.}$, the energy
$E$ and the circulation $\Gamma$. We have shown that equilibrium states of this system
have the form of Beltrami mean flows (linear vorticity-stream function relationship)
on which are superimposed Gaussian fluctuations. We have  shown that the maximization of entropy $S$
at fixed energy $E$, circulation $\Gamma$ and microscopic enstrophy
$\Gamma_2^{f.g.}$ (microcanonical ensemble) is equivalent to the maximization of grand
potential ${\cal S}=S-\alpha_2 \Gamma_2^{f.g.}$ at fixed energy $E$ and
circulation $\Gamma$ (grand microcanonical ensemble). These variational principles are also equivalent  to the minimization of
macroscopic enstrophy $\Gamma_2^{c.g.}$ at fixed energy $E$ and circulation $\Gamma$. This provides a justification of a minimum enstrophy  principle (selective decay) from statistical mechanics. In the analogy between 2D turbulence and 3D axisymmetric turbulence, the energy plays the role of the enstrophy.

In the present paper, we study more closely
the equilibrium states of axisymmetric flows and explore their stability. We show that all
critical points of macroscopic energy at fixed helicity and angular momentum
are {\it saddle points}, so that they are unstable
in a strict sense. Indeed, there is no minimum (macroscopic) energy
state at fixed helicity and angular momentum (either globally or
locally) because we can always decrease the energy by considering a
perturbation at smaller scales. This is reminiscent of the Richardson
energy cascade in 3D turbulence. Inversely, in 2D turbulence, there exists
 minimum enstrophy states that develop  at large scales (inverse cascade).
 Therefore, our system is intermediate
between 2D and 3D turbulence: there exists equilibrium states in the
form of coherent structures (that are solutions of a mean field
differential equation) like in 2D turbulence, but they are saddle
points of macroscopic energy and are expected to cascade towards smaller and
smaller scales like in 3D turbulence.  However, we give arguments
showing that saddle points  can be robust in practice and
play a role in the dynamics. Indeed, they are unstable only for some
particular (optimal) perturbations and can persist for a long time if
the system does not spontaneously generate these
perturbations. Therefore, these large-scale  coherent structures can play a role in
the dynamics and they have indeed been observed in experiments of von
K\'arm\'an flows \cite{cortet09}. In order to make this idea more precise, we have
explored their stability numerically using phenomenological
relaxation equations derived in
\cite{naso09}. We have found some domains of robustness in the parameter
space. In particular, the one cell structure is highly robust for
large values of the angular momentum $I>I_c$ and becomes weakly robust
for low values of the angular momentum. In that case, we expect a
phase transition (bifurcation) from the one-cell structure to the
two-cells structure. We have also found that the value of the critical
angular momentum $I_c$ changes depending whether we use relaxation
equations associated with a canonical (fixed temperature) or
microcanonical (fixed microscopic energy) description.  At low
temperatures $T$, we have evidence a new kind of ``ensembles
inequivalence'' characterizing the robustness of
saddle points with respect to random perturbations.

The paper is organized as follows: In Sec. \ref{SetUp}, we set-up
the various notations and hypotheses we are going to use. The
computation and characterization of equilibrium states is done in Sec.
\ref{Computation}. The stability analysis of these equilibrium states is
performed in Sec. \ref{Analytics} where we show analytically
 that all states are unstable with respect to
large wavenumber perturbations. We evidence a process of energy
condensation at small scales that is reminiscent of the Richardson
cascade. We explore numerically the robustness of the equilibria in
both  canonical and microcanonical ensembles in Sec.
\ref{Stability}. Our numerical method is probabilistic and rather
involved.  A
discussion of our results is done in Sec. \ref{Discussion} where a
bifurcation scenario relevant to the turbulent experimental von K\'arm\'an flow
is suggested.

\section{Theoretical set-up}
\label{SetUp}
\subsection{Hypotheses and Notations}

We consider a system with a cylindrical geometry enclosed in the
volume delimited above and below by surfaces $z=0$ and
$z=2h$, and radially by $0\le r \le R$. Like in Paper I, we consider
an axisymmetric Euler-Beltrami system characterized by a velocity
field ${\bf u}$,  with axisymmetric time averaged $\overline{\bf u}$.
 We furthermore assume that 
the only relevant invariants of the axisymmetric
Euler equations for our problem are the averaged
energy $E=\frac{1}{2}\int
\overline{{\bf u}^2}\, d{\bf r}$, the averaged helicity
$H=\int\overline{{\bf u}\cdot{\boldsymbol \omega}}\, d{\bf r}$ and the averaged
angular momentum $I=\int \overline{\sigma}\, d{\bf r}$ where
$\sigma=ru_\theta$.  We introduce the potential
vorticity $\xi=\omega_\theta/r$ and the stream function $\psi$ such
that $u_r = -r^{-1}\partial_z \psi $ and $u_z = r^{-1}\partial_r \psi
$. They are related to each other by the generalized Laplacian operator
\begin{equation}
\Delta_*\psi\equiv \frac{1}{r^2}\frac{\partial^2\psi}{\partial z^2}+\frac{1}{r}\frac{\partial}{\partial r}\left (\frac{1}{r}\frac{\partial\psi}{\partial r}\right )=-\xi.
\label{su1}
\end{equation}
In actual turbulent von K\'arm\'an experiments, we have been able to observe 
that the largest part of the kinetic energy is contained in the
toroidal motions. It is therefore natural, as a first elementary
step, to consider a model in which only toroidal fluctuations are
considered, and suppose that the fluctuations in the other (poloidal)
directions are simply frozen. With such an assumption, poloidal
vorticity fluctuations are allowed, but toroidal vorticity
fluctuations are excluded. We therefore only include a fraction of the
vorticity fluctuations, that presumably become predominant at
small scale, due to the existence of vortex stretching. As shown
below and in the next paper \cite{dubrulle09}, this simplification
however still allows for vortex stretching and energy cascades towards
smaller scales, and leads to predictions that are in good agreement with
experiments. Moreover, our hypotheses lead to a model that is
self-contained and analytically tractable. According to our
hypotheses, neither $\xi$ nor $\psi$ fluctuates in time:
$\xi=\overline{\xi}$ and $\psi=\overline{\psi}$. In that case, the
conserved quantities can be rewritten
\begin{eqnarray}
E^{f.g.} =  \displaystyle\frac{1}{2} \left\langle \overline{\xi}\psi\right\rangle  + \displaystyle\frac{1}{2} \left\langle \displaystyle\frac{\overline{\sigma^2}}{r^2}\right\rangle,
\label{dE}
\end{eqnarray}
\begin{eqnarray}
H = \left\langle \overline{\xi}\, \overline{\sigma}\right\rangle,
\label{dH}
\end{eqnarray}
\begin{eqnarray}
I = \left\langle \overline{\sigma}\right\rangle,
\label{dI}
\end{eqnarray}
where $\langle f \rangle$ denotes the spatial average \footnote{All integral constraints are normalized by the volume $V=2\pi R^2h$.}
\begin{equation}
\left\langle f \right\rangle \equiv \frac{1}{hR^2}\int_0^R \int_{0}^{2h} rdrdz\; f(r,z).
\label{prod_sc}
\end{equation}
The helicity and the angular momentum are {\it robust constraints} because they can be expressed in terms of coarse-grained quantities $\overline{\xi}$ and $\overline{\sigma}$. By
contrast, the energy is a {\it fragile constraint} because it cannot
be expressed in terms of coarse-grained quantities. Indeed, it
involves the fluctuations of angular momentum $\overline{\sigma^2}$. To emphasize that
point, we have introduced the notation $E\equiv E^{f.g.}$ to designate
the fine-grained (microscopic) energy.  Splitting $\sigma$ into a mean
part $\overline{\sigma}$ and a fluctuating part $\delta
\sigma$, we define the coarse-grained (macroscopic) energy by
\begin{eqnarray}
E^{c.g.}= \displaystyle\frac{1}{2} \left\langle \overline{\xi}\psi\right\rangle+ \displaystyle\frac{1}{2} \left\langle \displaystyle\frac{\overline{\sigma}^2}{r^2}\right\rangle. \label{E_cg}
\end{eqnarray}
Then, the energy contained in the fluctuations is simply
\begin{eqnarray}
E_{fluct}\equiv E^{f.g.}-E^{c.g.}= \displaystyle\frac{1}{2} \left\langle \displaystyle\frac{{\sigma}_2}{r^2}\right\rangle,
\label{fluct}
\end{eqnarray}
where
\begin{equation}
\sigma_2\equiv \overline{\sigma^2} - {\overline{\sigma}} ^2,
\end{equation}
is the local centered variance of angular momentum. We stress that the microscopic energy $E=E^{f.g.}$ is conserved while the macroscopic energy $E^{c.g.}$ is {\it not} conserved and is likely to decrease (see below).

\subsection{The two statistical ensembles and the selective decay principle}

In Paper I, we have developed a simplified thermodynamic approach of axisymmetric flows under the above-mentioned hypothesis. Let $\rho({\bf r},\eta)$ denote the PDF of $\sigma$ and let us recall the expression of the entropy
\begin{equation}
S\left\lbrack\rho\right\rbrack=-\int \rho\ln\rho\, d{\bf r}d\eta.
\end{equation}
We have proven the equivalence between the {\it microcanonical} ensemble
\begin{equation}
\max_{\rho,\overline{\xi}}\lbrace S[\rho]\, | \, E^{f.g.}, \, H,\, I, \, \int \rho d\eta=1  \rbrace,
\label{bes1}
\end{equation}
and the {\it canonical} ensemble
\begin{equation}
\max_{\rho,\overline{\xi}}\lbrace J[\rho]=S-\beta E^{f.g.}\, |  \, H,\, I, \, \int \rho d\eta=1  \rbrace.
\label{bes2}
\end{equation}
In each ensemble, the critical points are determined by the first order condition $\delta S-\beta\delta E^{f.g.}-\mu\delta H-\alpha \delta I=0$. The equilibrium distribution is Gaussian
\begin{equation}
\rho({\bf r},\eta)=\left (\frac{\beta}{2\pi r^2}\right )^{1/2}e^{-\frac{\beta}{2r^2}(\eta-\overline{\sigma})^2},
\label{vp1}
\end{equation}
the mean flow is a Beltrami state
\begin{eqnarray}
\overline{\sigma} = -\frac{\beta}{\mu}{\psi},
\label{vp2}
\end{eqnarray}
\begin{eqnarray}
\overline{\xi} = -\frac{\beta \overline{\sigma}}{\mu r^2} -\frac{\alpha}{\mu},
\label{vp3}
\end{eqnarray}
and the centered variance of angular momentum is
\begin{equation}
\sigma_2 = \displaystyle\frac{r^2}{\beta}.
\label{vp4}
\end{equation}
These equations determine {\it critical points} of the variational problems (\ref{bes1}) and (\ref{bes2}) that cancel the  first order variations of the thermodynamical potential. Clearly, (\ref{bes1}) and (\ref{bes2}) have the same critical points. Furthermore, it is shown in Paper I that (\ref{bes1}) and (\ref{bes2}) are equivalent for the {\it maximization} problem linked with the sign of the second order variations of the thermodynamical potential: a critical point determined by Eqs. (\ref{vp1})-(\ref{vp4}) is a maximum of $S$ at fixed microscopic energy, helicity and angular momentum  iff it is a maximum  of $J$ at fixed helicity and angular momentum. This equivalence  is not generic. We always have the implication (\ref{bes2}) $\Rightarrow$ (\ref{bes1})  but the reciprocal may be wrong. Here, the microcanonical and canonical ensembles are equivalent due to the quadratic nature of the microscopic energy $\sim\overline{\sigma^2}$. We note that, according to Eq. (\ref{vp4}), $\beta$ is positive. In the canonical ensemble, $\beta$ is prescribed. In the microcanonical ensemble, $\beta$ is a Lagrange multiplier that must be related to the energy $E=E^{f.g.}$. According to Eqs. (\ref{fluct}) and (\ref{vp4}),  we find that $\beta>0$ is determined by the condition
\begin{equation}
E^{f.g.}-E^{c.g.}=E_{fluct} =  \frac{1}{2\beta}.
\end{equation}
This relation shows that  $T=1/\beta$ plays the role of a temperature associated with the fluctuations of angular momentum \footnote{In particular, the temperature in 3D is positive contrary to the 2D case. In fact, in the present context, the fine-grained energy in 3D is the counterpart of the fine-grained enstrophy in 2D and the inverse temperature $\beta$ in 3D is the counterpart of the chemical potential $\alpha_2>0$ associated with the conservation of the fine-grained enstrophy in 2D (see Introduction).}. Finally, we have proven in Paper I that the two variational problems (\ref{bes1}) and (\ref{bes2}) are equivalent to
\begin{equation}
\max_{\overline{\sigma},\overline{\xi}}\lbrace J[\overline{\sigma},\overline{\xi}]=-\beta E^{c.g.}\, | \, H,\, I \rbrace,\label{res13}
\end{equation}
or equivalently
\begin{equation}
\min_{\overline{\sigma},\overline{\xi}}\lbrace E^{c.g.}[\overline{\sigma},\overline{\xi}]\, | \, H,\, I \rbrace,\label{res13b}
\end{equation}
in the sense that the solution of (\ref{bes1}) or (\ref{bes2})  is given by Eq. (\ref{vp1}) where $(\overline{\sigma},\overline{\xi})$ are the solutions of (\ref{res13}) or (\ref{res13b}). This justifies a {\it selective decay principle} from statistical mechanics. Indeed, it is often argued that an axisymmetric turbulent flow should evolve so as to minimize energy at fixed helicity and angular momentum. In general, this phenomenological principle is motivated by viscosity or other dissipative processes. In our approach, it is justified by the maximum entropy principle (\ref{bes1}) of statistical mechanics when a coarse-graining is introduced. In the sequel, we shall study  the maximization problem (\ref{res13b}) since it is simpler than (\ref{bes1}) or (\ref{bes2}), albeit equivalent.

{\it Remark:} although the variational problems (\ref{bes1}) and (\ref{bes2}) determining equilibrium states are equivalent, this does not mean that the relaxation equations associated with these variational problems are equivalent. To take an analogy, the Boltzmann (microcanonical) and the Kramers (canonical) equations have the same equilibrium states -the Maxwell distribution- but a  different dynamics. In the following, we will show that the equilibrium variational problems (\ref{bes1}) and (\ref{bes2}) have no solution. Indeed, there is no maximum of entropy at fixed $E^{f.g.}$, $H$ and $I$ and  no minimum of  free energy at fixed $H$ and $I$. All the critical points of (\ref{bes1}) and (\ref{bes2})  are saddle points of the thermodynamical potentials. Then, the idea is to consider the out-of-equilibrium problem, introduce relaxation equations and study the robustness of saddle points with respect to random perturbations. For what concerns the out-of-equilibrium problem, the microcanonical and canonical ensembles may be inequivalent. We will see that they are indeed inequivalent.

\section{Minimum energy states}
\label{Computation}

\subsection{Critical points}

In this section, we shall study the minimization problem
\begin{equation}
\min_{\overline{\sigma},\overline{\xi}}\lbrace E^{c.g.}[\overline{\sigma},\overline{\xi}]\, | \, H,\, I \rbrace.\label{res13c}
\end{equation}
The critical points of macroscopic energy at fixed helicity and angular momentum are determined by the condition
\begin{equation}
\delta E^{c.g.} +\mu\delta H+\alpha\delta I=0,
\end{equation}
where $\mu$ (helical potential) and $\alpha$ (chemical potential) are Lagrange multipliers.  Introducing the notations $B = -{1}/{\mu}$ and $C =
-{\alpha}/{\mu}$, the variations on
$\delta\overline{\xi}$ and $\delta\overline{\sigma}$ lead to
\begin{eqnarray}
\overline{\sigma} = B {\psi},
\label{bel1}
\end{eqnarray}
\begin{eqnarray}
\overline{\xi} = B\displaystyle\frac{\overline{\sigma}}{r^2} + C,
\label{bel2}
\end{eqnarray}
which are equivalent to Eqs. (\ref{vp2})-(\ref{vp3}) up to a change of notations. In the following, it will be convenient to work with the new field $\phi=\psi/r$. It is easy to check that
\begin{eqnarray}
\Delta_*\psi=\frac{1}{r}\left (\Delta\phi-\frac{\phi}{r^2}\right ),
\end{eqnarray}
where $\Delta$ is the usual Laplacian. Therefore, Eq. (\ref{su1}) becomes
\begin{eqnarray}
-\Delta\phi+\frac{\phi}{r^2}=r\overline{\xi},
\label{pz}
\end{eqnarray}
 and the previous equations can be rewritten
\begin{eqnarray}
\overline{\sigma}= Br\phi,
\end{eqnarray}
\begin{eqnarray}
\overline{\xi}= \frac{B^2}{r}\phi+C,
\end{eqnarray}
where $\phi$ is solution of
\begin{eqnarray}
-\Delta \phi+\frac{\phi}{r^2}=B^2\phi+Cr,
\label{phi}
\end{eqnarray}
with $\phi=0$ on the boundary. This is the fundamental differential
equation of the problem. Note that a particular solution of this differential equation
is
\begin{eqnarray}
\label{de19}
\phi_{part}=-\frac{C}{B^2}r,
\end{eqnarray}
but it does not satisfy the boundary conditions. Using Eqs. (\ref{dH}) and (\ref{dI}), the helicity and
the angular momentum are given by
\begin{eqnarray}
\label{de20}
H-CI=B^3\langle\phi^2\rangle,
\end{eqnarray}
\begin{eqnarray}
\label{de21}
I=B\langle\phi r\rangle.
\end{eqnarray}
These equations are relationships between $(B,C)$ and $(H,I)$.

{\it Remark:} we have not taken into account the conservation of circulation $\Gamma=\langle \xi\rangle$ because this would lead to a term $AB/r$ in the r.h.s. of Eq. (\ref{phi}) that diverges as $r\rightarrow 0$.

\subsection{The different solutions}
\label{sec_ds}

To construct the different  solutions of Eq. (\ref{phi}) and study their stability, we shall
follow the general procedure developed by Chavanis \& Sommeria \cite{jfm96} for the 2D Euler equation. We first introduce an eigenmode decomposition to compute all critical points of (\ref{res13c}). Then, we investigate their stability by determining whether they are (local) minima of macroscopic energy or saddle points.

\subsubsection{The eigenmodes}
\label{sec_e}

We first assume that
\begin{eqnarray}
\label{e1}
C=0.
\end{eqnarray}
In that case, the differential equation (\ref{phi}) becomes
\begin{eqnarray}
\label{e2}
 -\Delta\phi+\frac{\phi}{r^2}=B^2\phi,
\end{eqnarray}
with $\phi=0$ on the domain boundary. We introduce the eigenfunctions $\phi_{mn}$ of the operator ${\cal L}\equiv -\Delta+\frac{1}{r^2}$. They are defined by
\begin{eqnarray}
\label{e3}
{\cal L}\phi_{mn}\equiv-\Delta\phi_{mn}+\frac{\phi_{mn}}{r^2}=B_{mn}^2\phi_{mn},
\end{eqnarray}
with $\phi_{mn}=0$ on the domain boundary. It is easy to show that the eigenvalues $\Lambda_{mn}$ of ${\cal L}$ are positive (hence the notation $\Lambda_{mn}=B_{mn}^2$). Indeed, we have $\langle \phi_{mn}{\cal L}\phi_{mn}\rangle=\Lambda_{mn}\langle\phi_{mn}^2\rangle$ and $\langle \phi_{mn}{\cal L}\phi_{mn}\rangle=\langle (\nabla\phi_{mn})^2\rangle+\langle\frac{\phi_{mn}^2}{r^2}\rangle\ge 0$, which proves the result. It is also easy to show that the eigenfunctions are orthogonal with respect to the scalar product
\begin{equation}
\left\langle fg \right\rangle \equiv \frac{1}{hR^2}\int_0^R \int_{0}^{2h} rdrdz\; f(r,z) g(r,z).
\label{prod_sc1}
\end{equation}
Finally, we normalize them so that  $\langle\phi_{mn}\phi_{m'n'}\rangle=\delta_{mm'}\delta_{nn'}$.

The eigenvalues and eigenfunctions of the operator ${\cal L}$ can be determined analytically. The differential equation (\ref{e3}) can be rewritten
\begin{eqnarray}
\label{e4}
\frac{\partial^2\phi}{\partial r^2}+\frac{1}{r}\frac{\partial\phi}{\partial r}+\frac{\partial^2\phi}{\partial z^2}-\frac{\phi}{r^2}=-B^2\phi.
\end{eqnarray}
We look for solutions in the form $\phi(r,z)=f(r)g(z)$. This yields
\begin{eqnarray}
\label{e5}
\frac{f''}{f}+\frac{1}{r}\frac{f'}{f}-\frac{1}{r^2}+B^2=-\frac{g''}{g}\equiv \kappa^2,
\end{eqnarray}
where the sign of the constant has been chosen in order to satisfy the boundary condition $\phi=0$ in $z=0$ and $z=2h$. The differential equation for $g$ is readily solved and we obtain
\begin{eqnarray}
\label{e6}
g(z)=\sin(\kappa z),
\end{eqnarray}
with
\begin{eqnarray}
\label{e6n}
\kappa=n\frac{\pi}{2h},
\end{eqnarray}
where $n$ is a strictly positive integer. On the other hand, the differential equation for $f$ is
\begin{eqnarray}
\label{e7}
r^2 f''+r f'-r^2 (\kappa^2-B^2)f-f=0.
\end{eqnarray}
If we define $\lambda^2=B^2-\kappa^2$ and $x=\lambda r$, the foregoing equation can be rewritten
\begin{eqnarray}
\label{e8}
x^2 f''+x f'+x^2 f-f=0.
\end{eqnarray}
This is a Bessel equation whose solution is
\begin{eqnarray}
\label{e9}
f=J_1(x).
\end{eqnarray}
Now, the boundary condition $\phi(R)=0$ implies $f(\lambda R)=0$ so that
\begin{eqnarray}
\label{e11}
\lambda R=j_{1m},
\end{eqnarray}
where $j_{1m}$ is the $m$-th zero of Bessel function $J_1$.
In conclusion, the eigenvalues are
\begin{eqnarray}
\label{e13}
{B_{mn}}^2=\left (\frac{j_{1m}}{R}\right )^2+\left (\frac{n\pi}{2h}\right )^2,
\end{eqnarray}
and the eigenfunctions are
\begin{eqnarray}
\label{e10}
\phi_{mn}={\cal N}_{mn}\, J_1\left (\frac{j_{1m}r}{R}\right )\sin\left (\frac{n\pi z}{2h}\right ),
\end{eqnarray}
with the normalization constant
\begin{equation}
{\cal N}_{mn}=\sqrt{\frac{2}{ J_2^2(j_{1m})}}.
\end{equation}
The mode $(m,n)$ corresponds to $m$ cells in the $r$-direction and $n$
cells in the $z$-direction. We shall distinguish two kinds of modes,
according to their properties regarding the symmetry
${\cal R}$ with respect to the plane $z=h$. The {\it odd eigenmodes}
denoted $\phi'_{mn}$ are such that ${\cal R}\phi'_{mn}=-\phi'_{mn}$
and correspond to $n$ even. They have zero mean value in the $z$
direction ($\int_{0}^{2h} \phi'_{mn}\, dz =0$). For example, the mode
$(1,2)$ is a two-cells solution in the vertical direction. The {\it
even eigenmodes} denoted $\phi''_{mn}$ are such that ${\cal
R}\phi''_{mn}=\phi''_{mn}$ and correspond to $n$ odd. They have non zero
mean value in the vertical direction ($\int_{0}^{2h} \phi''_{mn}\, dz
\neq 0$). In particular, the mode $(1,1)$ is a one-cell solution.

Returning to Eq. (\ref{e2}), this differential equation has solutions only for quantized values of $B=B_{mn}$ (eigenvalues) and the corresponding solutions (eigenfunctions) are
\begin{eqnarray}
\label{e16}
\phi=\left (\frac{H}{B_{mn}^3}\right )^{1/2}\phi_{mn},
\end{eqnarray}
where we have used the helicity constraint (\ref{de20}) to determine the normalization constant. Note that Eq. (\ref{de20}) implies that $B_{mn}$ and $H$ have the same sign, so that the square root is always defined. Substituting this result in Eq. (\ref{de21}), and introducing the control parameter
\begin{eqnarray}
\label{e17}
\Lambda=\frac{I^2}{H},
\end{eqnarray}
we find that these solutions exist only for $\Lambda=\Lambda_{mn}$ with
\begin{eqnarray}
\label{e18}
\Lambda_{mn}=\frac{1}{B_{mn}}\langle\phi_{mn} r\rangle^2.
\end{eqnarray}
For the odd eigenmodes $\phi'_{mn}$, we have $\Lambda=0$ and for the even eigenmodes $\phi''_{mn}$, we have ${\Lambda''_{mn}}=\frac{1}{B''_{mn}}\langle\phi''_{mn} r\rangle^2$.

\subsubsection{The continuum}
\label{sec_c}

We now assume that $C\neq 0$ and define
\begin{eqnarray}
\label{c1}
\phi_B=\frac{\phi}{C}.
\end{eqnarray}
In that case, the fundamental differential equation (\ref{phi}) becomes
\begin{eqnarray}
\label{c2}
-\Delta\phi_B+\frac{\phi_B}{r^2}=B^2\phi_B+r,
\end{eqnarray}
with $\phi_B=0$ on the domain boundary.  We also assume that $B\neq B_{mn}$. In that case, Eq. (\ref{c2}) admits a unique solution that can be obtained by expanding $\phi_B$ on the eigenmodes. Using the identity
\begin{eqnarray}
f=\sum_{mn} \langle f \phi_{mn}\rangle \phi_{mn},
\end{eqnarray}
we get
\begin{eqnarray}
\label{c3}
\phi_B=\sum_{mn} \frac{\langle\phi''_{mn}r\rangle}{{B''_{mn}}^2-B^2}\phi''_{mn}.
\end{eqnarray}
Of course, $\phi_B$ can also be obtained by solving the differential equation (\ref{phi}) numerically. Note that this solution is even since only the even modes are ``excited''. Substituting Eq. (\ref{c1})  in  Eq. (\ref{de21}), we obtain
\begin{eqnarray}
\label{c4}
C=\frac{I}{B\langle\phi_B r\rangle}.
\end{eqnarray}
Then, substituting Eqs. (\ref{c1}) and (\ref{c4}) in Eq. (\ref{de20}), we get
\begin{eqnarray}
\label{c5}
\Lambda=\frac{B\langle\phi_B r\rangle^2}{\langle\phi_B r\rangle+B^2\langle\phi_B^2\rangle}.
\end{eqnarray}
This equation gives a relationship between $B$ and $\Lambda$. Then, $C$ is determined by Eq. (\ref{c4}).  These equations  can therefore  be viewed as the
equations of state of the system. They  determine the branch formed by the solutions of the continuum. Using
\begin{eqnarray}
\langle \phi_B r\rangle=\sum_{mn} \frac{\langle\phi''_{mn}r\rangle^2}{{B''_{mn}}^2-B^2},
\end{eqnarray}
and
\begin{eqnarray}
\langle \phi_B^2\rangle=\sum_{mn} \frac{\langle\phi''_{mn}r\rangle^2}{({B''_{mn}}^2-B^2)^2},
\end{eqnarray}
we obtain
\begin{eqnarray}
\langle\phi_B r\rangle+B^2\langle\phi_B^2\rangle=\sum_{mn} \frac{(B''_{mn})^2\langle\phi''_{mn}r\rangle^2}{({B''_{mn}}^2-B^2)^2}> 0.
\end{eqnarray}
This implies that $B$ is of the same sign as $\Lambda$, hence $H$. Furthermore, $\Lambda$ is an odd function of $B$. In the sequel, we shall consider only cases with $H\geq 0$, i.e. $B\geq 0$ and $\Lambda \geq 0$ for illustration and figures.

Note that Eq. (\ref{c5})  involves the important function
\begin{eqnarray}
\label{c6}
F(B)\equiv \langle\phi_B r\rangle = \sum_{mn} \frac{\langle\phi''_{mn}r\rangle^2}{{B''_{mn}}^2-B^2}.
\end{eqnarray}
For $\Lambda=0$, the inverse helical potential is $B=0$ or $B=B_*^{(n)}$ where $B_*^{(n)}$ is any zero of $F$, i.e.
\begin{eqnarray}
\label{c7}
F(B_*^{(n)})=\langle\phi_{B_*^{(n)}} r\rangle=0.
\end{eqnarray}
For simplicity, we shall call $B_*=B_*^{(1)}$ the first zero of $F$. This first zero is always between the
first and the second even eigenmodes (see Appendix A). Its location with
respect to the first odd eigenmode $B'_{12}$ depends on the aspect
ratio of the cylinder:
for $h/R>0.53$, we have $B_*>B'_{12}$ (case L-for Large aspect ratio) while for $h/R< 0.53$,
$B_*< B'_{12}$ (case S-for Small aspect ratio).

\subsubsection{The mixed solutions}
\label{sec_m}

We now consider the case where $C\neq 0$ and $B=B_{mn}$. For $B=B''_{mn}$, we recover the eigenfunction  $\phi_{mn}''$  as a limit case. Therefore, the even
eigenmodes are limit points of the main branch. On the other hand, for
$B=B'_{mn}$, the solution of Eq. (\ref{c2}) is not unique. Indeed, we can
always add to the solution (\ref{c3}) an  eigenmode
$\chi_{mn}\phi_{mn}'$. This leads to the mixed solution
\begin{eqnarray}
\label{m1}
\phi_M=\sum_{m'n'}\frac{\langle\phi''_{m'n'}r\rangle}{(B''_{m'n'})^2-(B'_{mn})^2}\phi''_{m'n'} +\chi_{mn}\phi'_{mn}.
\end{eqnarray}
The ``proportion''  $\chi_{mn}$  of the eigenmode  present in the mixed solution is determined by the control parameter $\Lambda$. Taking the norm of $\phi_M$ and its scalar product with $r$, we get
\begin{eqnarray}
\langle \phi_M^2\rangle&=&\chi_{mn}^2+\sum_{m'n'} \frac{\langle\phi''_{m'n'}r\rangle^2}{((B''_{m'n'})^2-(B'_{mn})^2)^2},\nonumber\\
&=& \chi_{mn}^2+
\langle \phi_B^2\rangle,\nonumber\\
\langle \phi_M r\rangle&=&\sum_{m'n'} \frac{\langle\phi''_{m'n'}r\rangle^2}{(B''_{m'n'})^2-(B'_{mn})^2}=
\langle \phi_B r\rangle.
\end{eqnarray}
Substituting these results in Eqs. (\ref{de20}) and (\ref{de21}), we find that $\chi_{mn}$ is
determined by $\Lambda$ according to
\begin{equation}
\Lambda=\frac{B\langle\phi_B r\rangle^2}{\langle\phi_B r\rangle+B^2(\chi_{mn}^2+\langle\phi_B^2\rangle)},
\label{detxi}
\end{equation}
with $B=B'_{mn}$. These mixed solutions exist in the range $0\le\Lambda\le\Lambda(B'_{mn})$ and they form a plateau at constant $B=B'_{mn}$. For $\chi_{mn}\rightarrow +\infty$, we recover the odd eigenmode $\phi'_{mn}$ at $\Lambda=0$ and for  $\chi_{mn}=0$, the plateau connects the branch of continuum solutions. The mixed solutions are therefore symmetry breaking solutions. They  can be seen as a mixture of a continuum solution and an eigenmode solution, like in situations with different phase coexistence.

\subsection{The helical potential curve}

\begin{figure}[!htb]
\begin{minipage}{0.99\columnwidth}
\includegraphics[width=0.99\columnwidth]{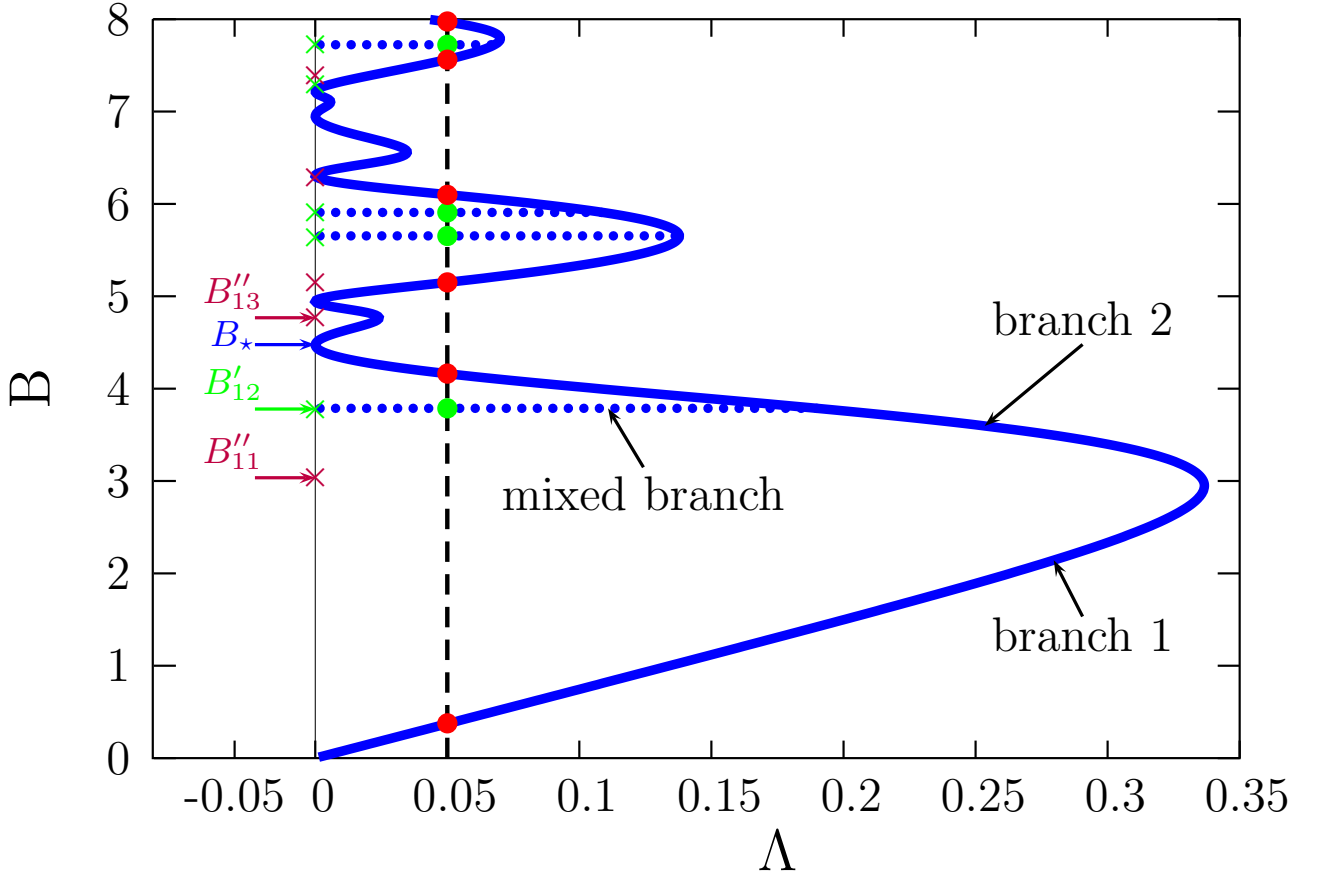}
\end{minipage}
\caption{Top: $B$ as a function of $\Lambda$ for case L (we have taken $R=1.4$ and  $h=1.2$). For a given value of $\Lambda$ (we have taken $\Lambda=0.05$), the solutions of the continuum  are denoted by red circles and the  mixed solutions by green circles. The mixed solution branches are drawn using dotted lines. One observes multiplicity of solutions: at given
$\Lambda$ correspond several solutions with different $B$.}
\label{BlamL}
\end{figure}

\begin{figure}
\begin{minipage}{0.99\columnwidth}
\includegraphics[width=0.99\columnwidth]{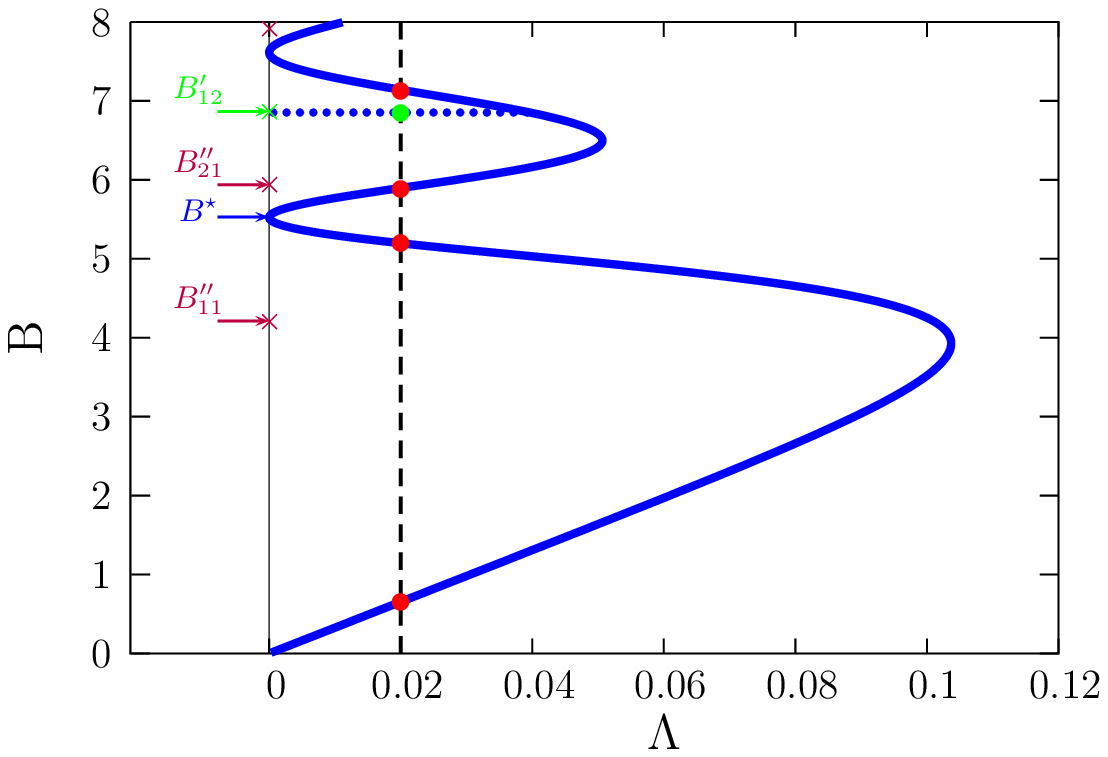}
\end{minipage}
\caption{Top: $B$ as a function of $\Lambda$ for case S (here $R=1.4$ and $h=0.5$). The solutions of the continuum  are denoted by red circles and the  mixed solutions by green circles. The mixed solution branches  are drawn using dotted lines.}\label{BlamS}
\end{figure}

In this section, we  plot $B$ as a function of $\Lambda$. For given $I\neq 0$, this curve determines the inverse helical potential $1/\mu$ as a function of the inverse helicity $1/H$ (conjugate variables). It is represented in Figs. \ref{BlamL} and \ref{BlamS} for the cases L and S respectively. One sees that, for a given value of the control parameter
$\Lambda$, there exists multiple solutions with different values of $B$.
We will see in Sec. \ref{coarsegrainedenergy}.  that, for a given value of $\Lambda$,  the macroscopic energy $E^{c.g.}$ decreases as $B$  increases. Therefore, low values of $B$ correspond to high energies states and high values of $B$ correspond to low energies states.

\begin{figure}[!htb]
    \centering
        \includegraphics[width=0.2\columnwidth,clip=true]{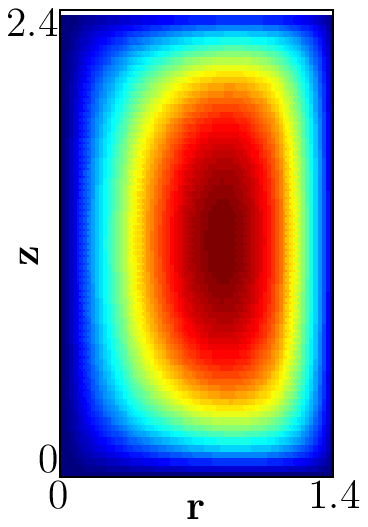}
		\includegraphics[width=0.2\columnwidth,clip=true]{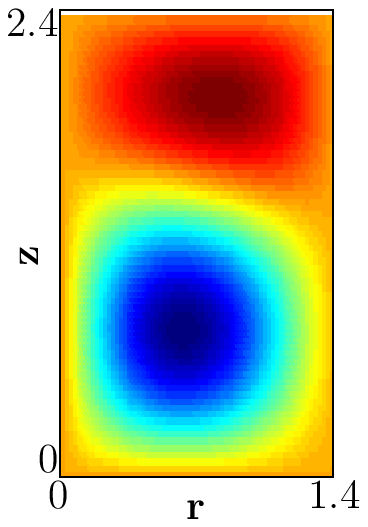}
		\includegraphics[width=0.2\columnwidth,clip=true]{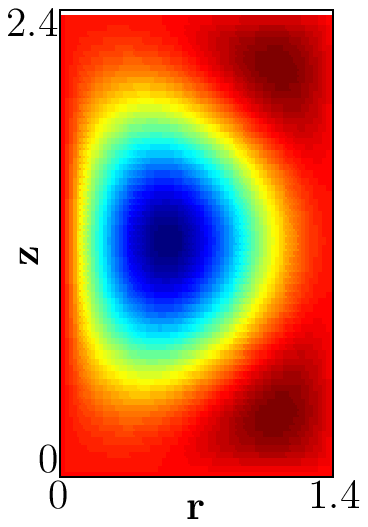}
		\includegraphics[width=0.2\columnwidth,clip=true]{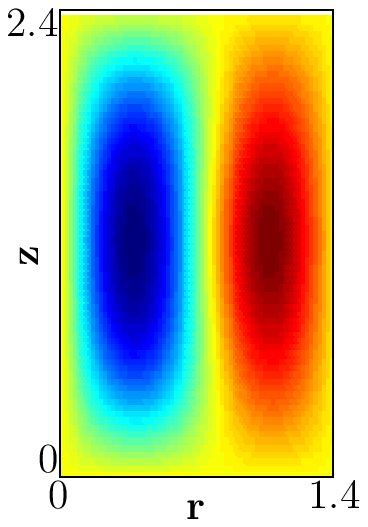}

\caption{Example of stream function $\phi$ of the four first solutions  for $\Lambda = 0.05$. From left to right:
$B=0.3767$ (direct monopole), $B=3.7874$ (vertical dipole), $B=4.1633$ (reversed monopole) and $B=5.15$. Increasing values from blue to red. By convention, we call direct (resp. reversed) monopole the one-cell solution with maximal (resp. minimal) inner stream function-see above. For simplicity, we show at each point only one solution, corresponding to a given sign of $I$. The solution corresponding to opposite sign of $I$ can be found by a change $\phi\to -\phi$. }
\label{fig:sel_ex2}
\end{figure}

\begin{figure}[!htb]
    \centering
        \includegraphics[width=0.2\columnwidth]{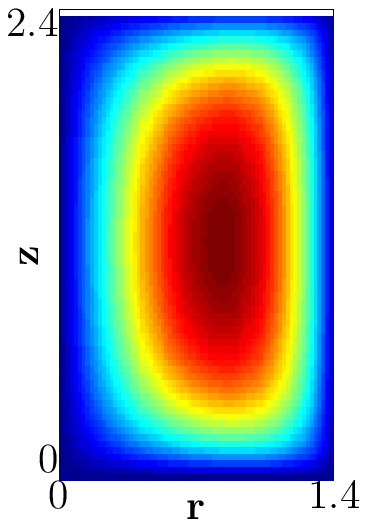}
		\includegraphics[width=0.2\columnwidth]{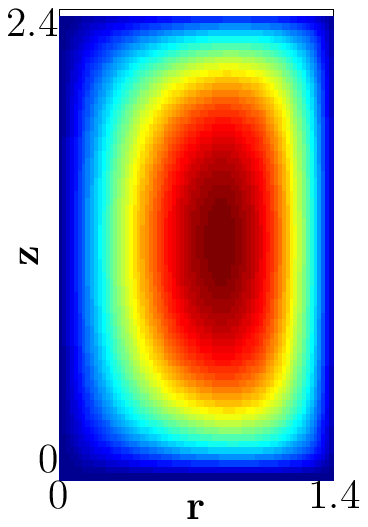}
		\includegraphics[width=0.2\columnwidth]{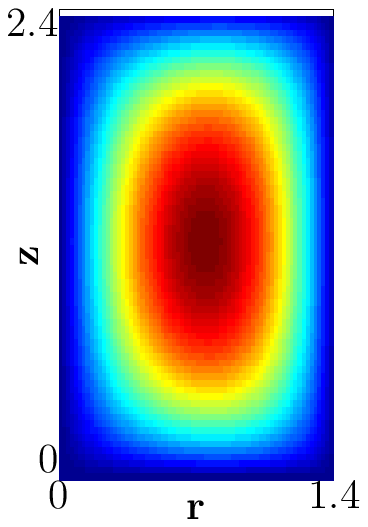}\\
		\vspace{0.5cm}
		\includegraphics[width=0.2\columnwidth]{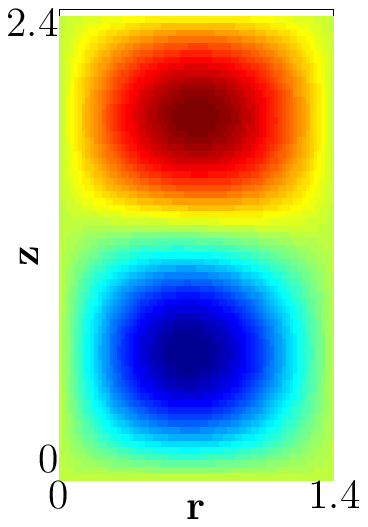}
		\includegraphics[width=0.2\columnwidth]{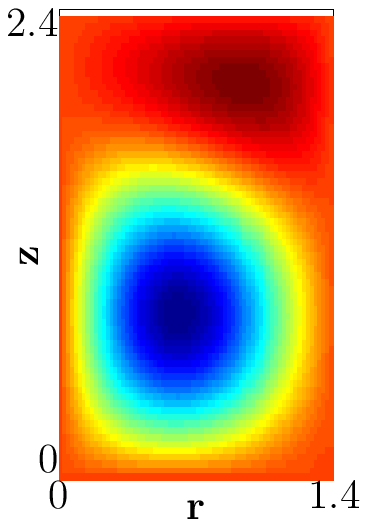}
		\includegraphics[width=0.2\columnwidth]{blank1.eps}\\
		\vspace{0.5cm}
		\includegraphics[width=0.2\columnwidth]{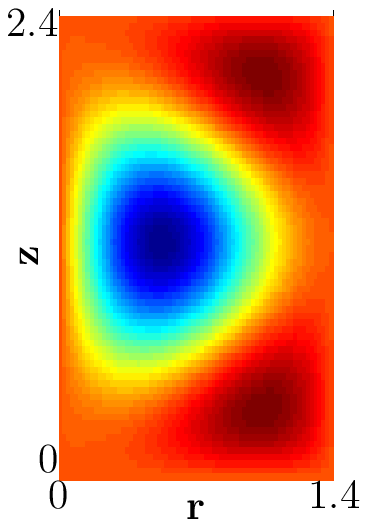}
		\includegraphics[width=0.2\columnwidth]{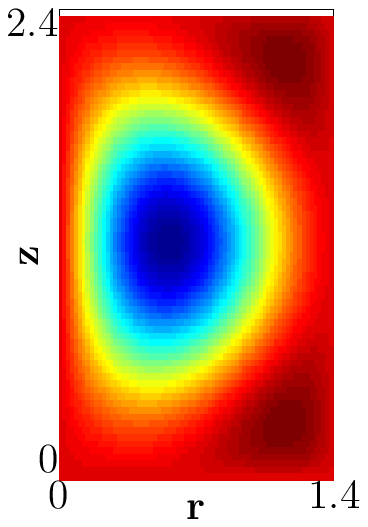}
		\includegraphics[width=0.2\columnwidth]{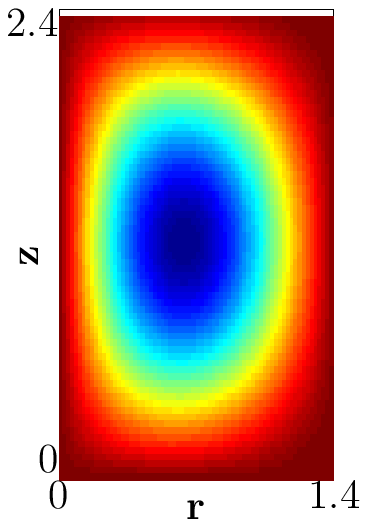}

 \caption{Stream function $\phi$ along the three branches of solution at $\Lambda = 0.004$, $0.1$ and
 $0.3$
 from left to right for each branch. Top= branch 1, direct monopole; Middle: mixed branch (vertical dipole); Bottom: branch 2,reversed monopole. For simplicity, we show at each point only one solution, corresponding to a given sign of $I$. The solution corresponding to opposite sign of $I$ can be found by a change $\phi\to -\phi$.}
\label{fig:branches}
\end{figure}

{\it Case L:} In this case $B_*>B'_{12}$ and the curve $B(\Lambda)$ looks typically
like in Fig. \ref{BlamL}. For a given value of $\Lambda$, we have different solutions
as represented in Fig. \ref{fig:sel_ex2}. The highest energy
solution is a one cell solution (continuum branch), that we choose to call "direct monopole". The second one is
a two vertical cells solution (mixed branch). The cells are symmetric for $\Lambda=0$ but one of the two cells grows for increasing $\Lambda$.
 The third highest energy solution is another
one-cell solution (continuum branch) rotating in a direction opposite
to that of the highest energy solution. We therefore call it a "reversed monopole". We call
these three respective branches of solutions ``branch 1'' and ``branch 2''
for the continuum solutions, and ``mixed branch'' for the mixed
solutions. The branches 1 and 2 connect each other at
$\Lambda(B''_{11})$, the location of the first even eigenmode. A typical sequence of variation of the stream function with
increasing $\Lambda$ on these three branches is given in Fig. \ref{fig:branches}. One sees that, as we increase $\Lambda$ on the mixed branch,  the two
cells solution, with a mixing layer at $z=h$ continuously
 transforms itself
into a one cell solution, via a continuous shift of the mixing layer
towards the vertical boundary.

\begin{figure}[h]
\begin{minipage}{0.99\columnwidth}
 \includegraphics[width=0.24\columnwidth]{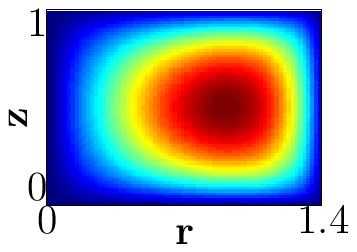}
 \includegraphics[width=0.24\columnwidth]{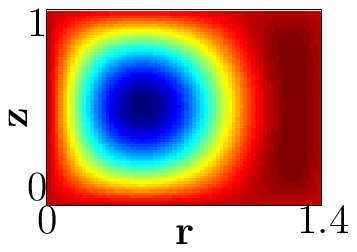}
\includegraphics[width=0.24\columnwidth]
{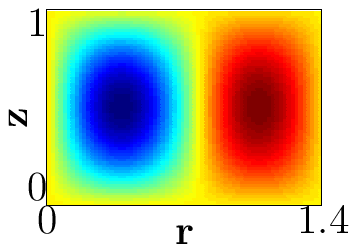}
\includegraphics[width=0.24\columnwidth]
{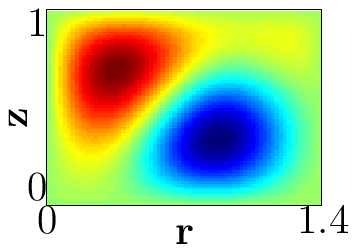}
\end{minipage}
\caption{Stream function $\phi$ of the four first solutions at $\Lambda = 0.02$ for case S. From left to right:
$B=0.6567$ (direct monopole), $B=5.2033$ (reversed monopole), $B=5.8967$ and $B=6.8534$. For simplicity, we show at each point only one solution, corresponding to a given sign of $I$. The solution corresponding to opposite sign of $I$ can be found by a change $\phi\to -\phi$.}
\label{Bifu04}
\end{figure}

{\it Case S:} In this case $B_*< B'_{12}$ and the curve $B(\Lambda)$ looks typically
like in Fig. \ref{BlamS}.
The highest energy solution is a one cell  solution (continuum branch), (direct monopole). The second solution is another one-cell  solution (continuum branch ) rotating in the opposite direction (reversed monopole). The third solution  is a two horizontal cells solutions (continuum branch). Some stream functions are represented in Fig. \ref{Bifu04}.

{\it Remark:} there is a maximum value of $\Lambda_{max}$ above which there is no critical point of energy at fixed helicity and angular momentum. In that case, the system is expected to cascade towards smaller and smaller scales since there is no possibility to be blocked in a ``saddle point''. This is a bit similar to the Antonov instability  in stellar dynamics due to the absence of critical point of entropy at fixed mass and energy below a critical value of energy \cite{antonov,lbw,paddy,ijmpb}. In that case, the system is expected to collapse (gravothermal catastrophe). It is not yet clear whether a similar process can be achieved in experiments of turbulent axisymmetric flows. For $\Lambda>\Lambda_{max}$, the system could become non-axisymmetric ruling out the theoretical analysis.

\subsection{The coarse-grained energy}
\label{coarsegrainedenergy}

\begin{figure}[!htb]
   \centering
        \includegraphics[width=0.45\columnwidth]{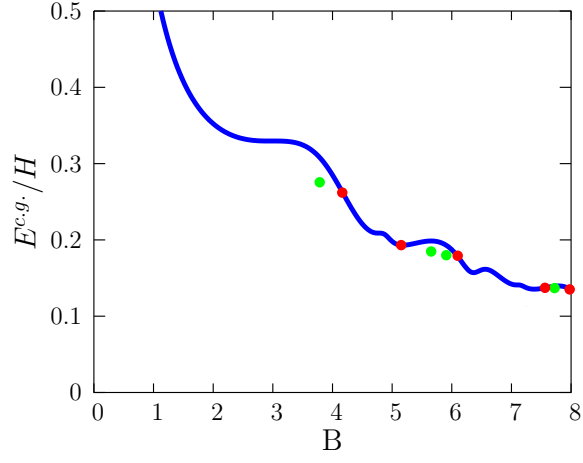}

\caption{Macroscopic energy $E^{c.g}/H$ as a function of $B$ for case L. The energy of the even eigenmodes  are denoted by red circles and the  energy of the odd eigenmodes by green circles.}
\label{fig:sel_ex}
\end{figure}

\begin{figure}
   \centering
        \includegraphics[width=0.45\columnwidth]{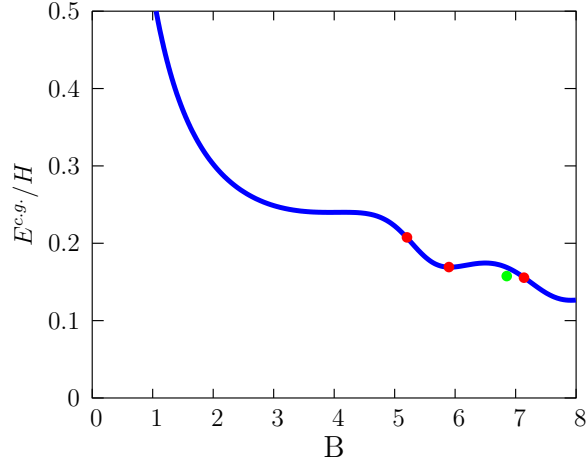}

\caption{Macroscopic energy $E^{c.g}/H$  as a function of $B$ for case S. The energy of the even eigenmodes  are denoted by red circles and the  energy of the odd eigenmodes by green circles.}\label{EvsB04}
\end{figure}

In the previous section, we have found several solutions  with
different values of $B$ for each value
of the control parameter $\Lambda< \Lambda_{max}$. According to the variational principle (\ref{res13c}), we should select the solution with the minimum macroscopic energy. Combining
Eqs. (\ref{dH}), (\ref{dI}), (\ref{E_cg}), (\ref{bel1}) and (\ref{bel2}), we obtain the relation
\begin{eqnarray}
\label{de22}
E^{c.g}=\frac{1}{B}\left (H-\frac{1}{2}CI\right ).
\end{eqnarray}
For the eigenmodes ($C=0$), we find that
\begin{eqnarray}
\label{na1}
\frac{E^{c.g}}{H}=\frac{1}{B_{mn}}.
\end{eqnarray}
Let us consider the odd eigenmodes $\phi'_{mn}$ that exist for $\Lambda=0$ only. They are in competition with each other. We see
that there is no minimum energy state since the energy decreases when
$(m,n)$ increase, i.e. when the eigenmodes develop  smaller and
smaller scales. Therefore, the minimum energy state corresponds to the
structure concentrated at the smallest accessible scale. For the solutions of the continuum, using Eq. (\ref{c4}), the macroscopic energy is
\begin{eqnarray}
\label{na2}
\frac{E^{c.g}}{H}=\frac{1}{B} \left (1- \frac{\Lambda}{2B\langle\phi_B r\rangle}\right ).
\end{eqnarray}
We can easily plot it as a function of $B$ (see Figs. \ref{fig:sel_ex} and \ref{EvsB04}). Combining Figs. \ref{BlamL},  \ref{BlamS},  \ref{fig:sel_ex} and \ref{EvsB04},  we see that, for a given value of $\Lambda$, the solution with the smallest macroscopic energy corresponds to the highest $B$, i.e. to small-scale structures. This is in complete
opposition to what happens in  2D turbulence. In that case, the
counterpart of the macroscopic energy $E^{c.g.}$ is the macroscopic enstrophy $\Gamma_2^{c.g.}$ and
the minimum enstrophy state corresponds to structures spreading at the largest
scale. Strikingly, the bifurcation diagram in 2D turbulence \cite{jfm96} is reversed with respect to the present one.

In conclusion, there is no global minimum of macroscopic energy at
fixed helicity and angular momentum. We can always
decrease the macroscopic energy by considering
structures at smaller and smaller scales. Since (\ref{bes1}),
(\ref{bes2}) and (\ref{res13b}) are equivalent, we also conclude that
there is no global maximum of entropy at fixed microscopic energy,
helicity and angular momentum. We may note a similar fact in
astrophysics. It is well-known that a stellar system has no global
entropy maximum at fixed mass and energy
\cite{antonov,lbw,paddy,ijmpb}. This is associated to gravitational
collapse (called the gravothermal catastrophe in the microcanonical
ensemble) leading to the formation of binary stars. However, in the
astrophysical problem, there exists local entropy maxima (metastable
states) at fixed mass and energy if the energy is sufficiently high
(above the Antonov energy).  Similarly, we could
investigate the existence of metastable states in the present
problem. However, we will show in Sec. \ref{Analytics} that there is
no local minimum of macroscopic energy at fixed helicity and angular
momentum. All the critical points (\ref{bel1})-(\ref{bel2}) of the
variational problem (\ref{res13c}) are saddle points!

\subsection{Chemical potential curve}

\begin{figure}[!htb]
   \centering
        \includegraphics[width=0.55\columnwidth]{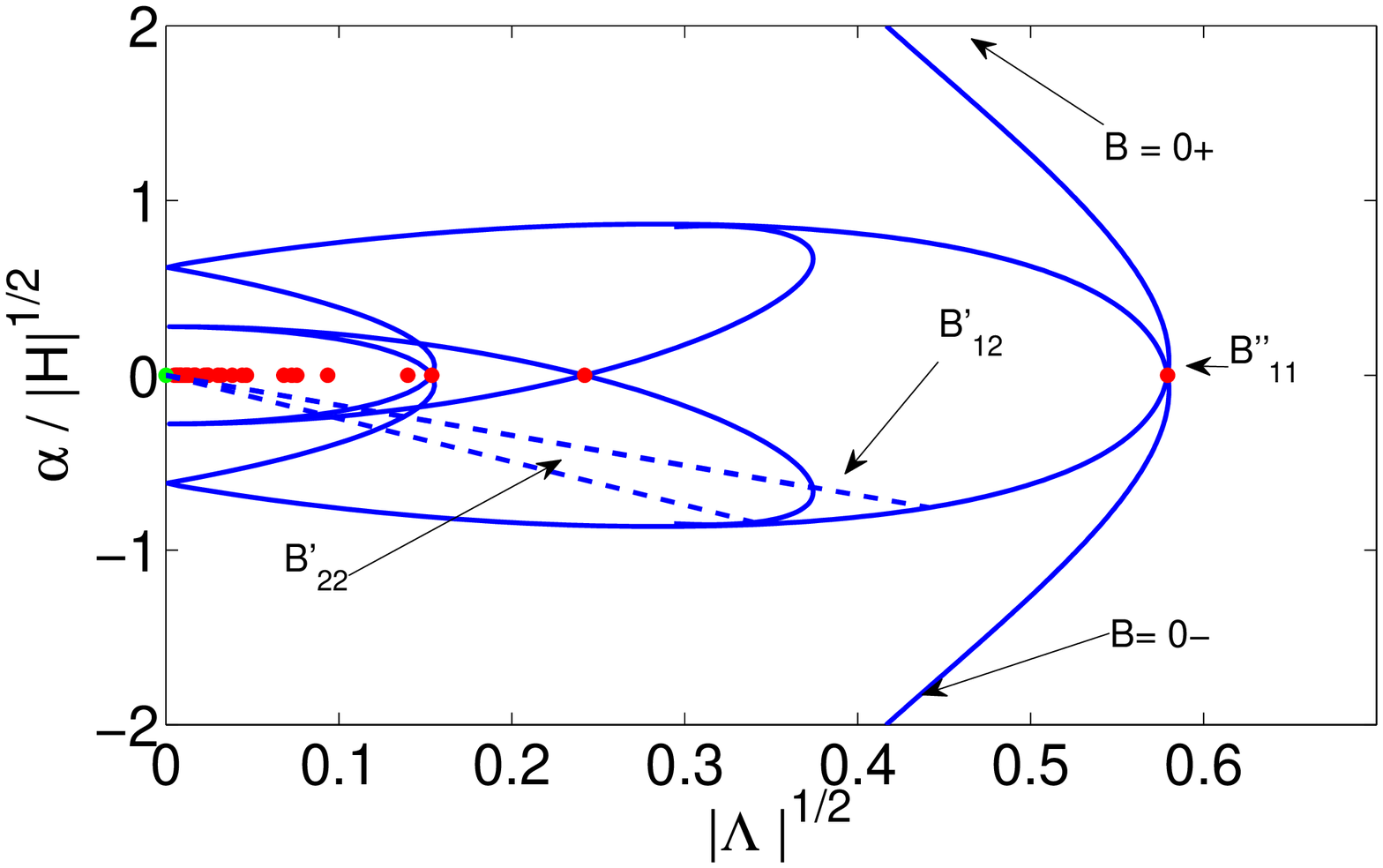}

\caption{Chemical potential versus angular momentum in case L. The chemical potential of the even eigenmodes  are denoted by red circles and the chemical potential of the odd eigenmodes by green circles. The mixed solution branches  are drawn using dotted lines.}
\label{CsursqrtH-caseL}
\end{figure}

\begin{figure}[!htb]
   \centering
        \includegraphics[width=0.55\columnwidth]{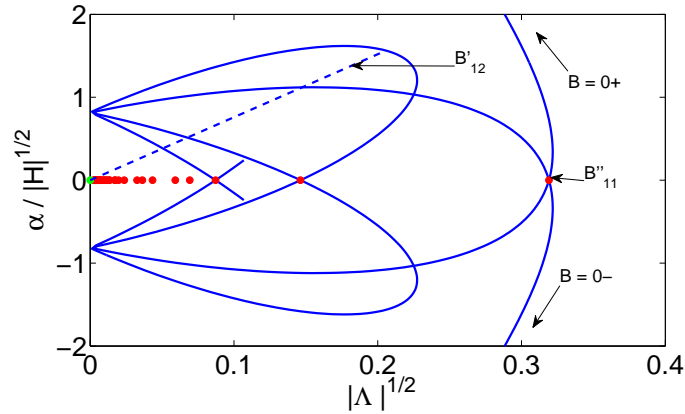}

\caption{Chemical potential versus angular momentum in case S. The chemical potential of the even eigenmodes  are denoted by red circles and the chemical potential of the odd eigenmodes by green circles. The mixed solution branches  are drawn using dotted lines.}
\label{CsursqrtH-caseS}
\end{figure}

In our system, the chemical potential is $\alpha=C/B$. For given $H$, we have to plot $\alpha$ as a function of $I$ (conjugate variables). The chemical potential is zero for the eigenmodes. Using the equation of state (\ref{c4}), we can express $\alpha$ for the continuum solutions as
\begin{eqnarray}
\label{chemical}
\frac{\alpha}{\sqrt{\vert H\vert}}&=&\frac{\sqrt{\vert\Lambda\vert}}{B^2\langle\phi_B r\rangle},
\end{eqnarray}
where $\Lambda$ is expressed as a function of $B$ by Eq. (\ref{c5}). Therefore, Eq. (\ref{chemical}) gives $\alpha/\sqrt{|H|}$ as a function of $B$. Eliminating $B$ between Eqs. (\ref{chemical}) and (\ref{c5}), we obtain $\alpha/\sqrt{|H|}$ as a function of $\Lambda$ for the continuum. For the mixed solutions, we have
\begin{eqnarray}
\label{chemicalM}
\frac{\alpha}{\sqrt{\vert H\vert}}&=&\frac{\sqrt{\vert\Lambda\vert}}{(B'_{mn})^2\langle\phi_{B'_{mn}} r\rangle},
\end{eqnarray}
corresponding to straight lines as a function of $\sqrt{\vert\Lambda\vert}$.  The chemical potential curve $\alpha/\sqrt{|H|}$ as a function of $\sqrt{\vert\Lambda\vert}$ is represented in Fig. \ref{CsursqrtH-caseL} for case L and in Fig. \ref{CsursqrtH-caseS} for case S. For fixed $H$, this gives $\alpha$ as a function of $I$.

 \subsection{Caloric curve }

If we come back to the initial variational problem (\ref{bes1}), the caloric curve should give $\beta$ as a function of the microscopic energy $E=E^{f.g.}$ (conjugate variables) for fixed values of $H$ and $I$. Now, the temperature is determined by the expression
\begin{equation}
\label{et}
E^{f.g.}=E^{c.g.}+\frac{1}{2\beta}.
\end{equation}
For given $H$ and $I$, we can determine the {\it discrete} values of $B_{(n)}$ and the corresponding {\it discrete} values of $E^{c.g.}_{(n)}$ as explained previously. Then, for each discrete value, the temperature is related to the energy by Eq. (\ref{et}). Therefore, the mean flow (Beltrami state) is fully determined by $H$ and $I$ and, for a given mean flow, the variance of the fluctuations (temperature) is determined by the energy $E=E^{f.g.}$ according to
\begin{equation}
\frac{1}{2}T=E-E^{c.g.}_{(n)}.
\end{equation}
In conclusion, the caloric curve $T(E)$, or more properly the series of equilibria, is formed by a  a discrete number of straight lines with value at the origin $-E^{c.g.}_{(n)}$ and with constant specific heats $C=dE/dT=1/2$. The specific heat is positive  since the microcanonical and canonical ensembles are equivalent in our problem.

\subsection{Stability analysis}
\label{Analytics}

In this section, we prove that the critical points of macroscopic energy at fixed helicity and angular momentum are all saddle points. A critical point of macroscopic energy at fixed helicity and angular momentum is a minimum (resp. maximum) iff the second order variations
\begin{eqnarray}
\label{sa1}
\delta^2F&\equiv& \delta^2 E^{c.g.}-\frac{\delta^2 H}{B}\nonumber\\
&=&\int \left(\frac{1}{2}r\delta\xi\delta\phi+ \frac{(\delta\sigma)^2}{2r^2}-\frac{1}{B}\delta\xi\delta\sigma\right)\, rdrdz,
\end{eqnarray}
are definite positive (resp. definite negative) for all perturbations that conserve helicity and angular momentum at first order, i.e. $\delta H=\langle\overline{\xi}\delta\sigma\rangle+\langle\overline{\sigma}\delta\xi\rangle=0$ and $\delta I=\langle\delta\sigma\rangle=0$. Adapting the procedure of Chavanis \& Sommeria \cite{jfm96} to the present context, we shall determine sufficient conditions of {\it instability}.

(i) Let us prove that there is no local maximum of macroscopic energy at fixed angular momentum and helicity.  Consider first the even solutions, including the continuum solutions and the even eigenmodes. We choose a  perturbation such that $\delta\sigma$ is odd and $\delta\xi=\delta\phi=0$. For symmetry reason, this perturbation  does not change $I$ nor $H$ at first order. On the other hand, for this perturbation $\delta^2 F=\int \frac{(\delta\sigma)^2}{2r^2}\, rdrdz>0$. Consider now the odd eigenmodes. We choose a  perturbation of the form $\delta\xi=\delta\phi=0$ and  $\delta\sigma=B_*r\phi_{B_*}$, where $\phi_{B_*}$ is the first continuum solution such that $\langle r\phi_{B_*}\rangle=0$. For this perturbation, we have  $\langle\delta\sigma\rangle=B_*\langle r\phi_{B_*}\rangle=0$, $\langle\overline{\sigma}\delta\xi\rangle=0$ and  $\langle\overline{\xi}\delta\sigma\rangle=(B'_{mn})^2B_*\langle \phi_{B_*}\phi'_{mn}\rangle=0$  since $\phi'_{mn}$ is orthogonal to $\phi_{B_*}$. Therefore, this perturbation does not change the helicity and the angular momentum at first order. On the other hand, for this perturbation $\delta^2 F=\int \frac{(\delta\sigma)^2}{2r^2}\, rdrdz>0$. As a result, the critical points of macroscopic energy at fixed helicity and angular momentum cannot be energy maxima since we can always find particular perturbations that increase the energy while conserving the constraints.

(ii) Let us prove that there is no local minimum of macroscopic energy at fixed angular momentum and helicity. To that purpose, we consider perturbations of the form $\delta \sigma=r\phi'_{MN}$ and $\delta\xi=B'_{MN}\phi'_{MN}/r$. The corresponding stream function is $\delta\phi=\phi'_{MN}/B'_{MN}$.
Consider first the even solutions, including the continuum solutions and the even eigenmodes.
In that case, we have  $\langle\delta\sigma\rangle=\langle r\phi'_{MN}\rangle=0$,  $\langle\overline{\xi}\delta\sigma\rangle=C B^2\langle \phi_B\phi'_{MN}\rangle+C\langle r\phi'_{MN}\rangle=0$  and $\langle\overline{\sigma}\delta\xi\rangle=C B B'_{MN}\langle \phi_B\phi'_{MN}\rangle=0$ since $\phi'_{MN}$ is orthogonal to $\phi_B$. The preceding relations remain valid for the odd eigenmodes $(m,n)$ provided that  $(M,N)\neq (m,n)$.
Therefore, these perturbations do not change the helicity and the angular momentum at first order. On the other hand, for these perturbations, we have
\begin{eqnarray}
\label{sa2}
\delta^2 F=1-\frac{B'_{MN}}{B}.
\end{eqnarray}
Thus, for given $B$ and $(M,N)$ sufficiently large \footnote{This automatically ensures that $(M,N)\neq (m,n)$ if the critical point is an odd eigenmode.}, i.e. $B'_{MN}>B$,  we have $\delta^2 F<0$. As a result, the critical points of macroscopic energy at fixed helicity and angular momentum cannot be energy minima since we can always find particular perturbations that decrease the energy while conserving the constraints.

In conclusion, the critical points of macroscopic energy at fixed helicity and angular momentum  are saddle points since we can find perturbations  making $\delta^2 F$ positive and perturbations   making $\delta^2 F$ negative. This analysis shows that all Beltrami solutions are unstable. However,  saddle points may be characterized by very long lifetimes as long as the system does not explore dangerous perturbations that destabilize them. This motivates the numerical stability analysis of Sec. \ref{Stability}.

{\it Remark:} Let us consider the odd eigenmode $(1,2)$. We have seen that it can be destabilized by a perturbation $\phi'_{14}$ or by a perturbation $\phi'_{MN}$ at smaller scale. Let us now consider the effect of a perturbation of the form $\delta\xi=1+B^2_* \phi_{B_*}/r$ and $\delta\sigma=B_*r\phi_{B_*}$, where $\phi_{B_*}$ is the first continuum mode such that $\langle r\phi_{B_*}\rangle=0$.  The corresponding stream
function is $\delta\phi=\phi_{B_*}$. For this perturbation, we have  $\langle\delta\sigma\rangle=B_*\langle r\phi_{B_*}\rangle=0$,  $\langle\overline{\xi}\delta\sigma\rangle=(B'_{12})^2B_*\langle \phi_{B_*}\phi'_{12}\rangle=0$  and $\langle\overline{\sigma}\delta\xi\rangle=B'_{12}\langle r\phi'_{12}\rangle+B'_{12}B_*^2\langle \phi'_{12}\phi_{B_*}\rangle=0$ since $\phi'_{12}$ is orthogonal to $\phi_{B_*}$. Therefore, this perturbation does not change the helicity and the angular momentum at first order. For this perturbation, we have in addition
\begin{eqnarray}
\label{sa3}
\delta^2 F=B_*^2\left (1-\frac{B_*}{B'_{12}}\right ) \langle\phi_{B_*}^2\rangle.
\end{eqnarray}
This quantity is negative when $B'_{12}< B_*$ corresponding to case L. This implies that the eigenmode $(1,2)$ is also destabilized by the perturbation  $\delta\phi=\phi_{B_*}$ which is at larger scale than the perturbations $\phi'_{14}$.

\section{Numerical stability analysis: robustness of saddle points}
\label{Stability}

The stability analysis performed in Sec. \ref{Analytics} has shown
that all the critical points of entropy at fixed microscopic energy,
helicity and angular momentum are saddle points. We shall now
investigate their robustness by using the relaxation equations derived
in Paper I (for a review of relaxation equations in
the context of 2D hydrodynamics, see \cite{revuerelax}).
These relaxation equations can serve as numerical
algorithms to compute maximum entropy states or minimum energy states
with relevant constraints. Their study is interesting in its own right
since these equations constitute non trivial dynamical systems leading
to rich bifurcations. Although these relaxation equations do not
provide a parametrization of turbulence (we have no rigorous argument
for that), they may however give an idea of the true dynamical
evolution of the flow. In that respect, it would be interesting to
compare these relaxation equations with Navier-Stokes
simulations. This will, however, not be attempted in the present
paper.

By construction, the relaxation equations monotonically increase entropy, or decrease energy, with relevant constraints. Different
generic evolutions are possible: (i) they can relax towards a fully
stable state (global maximum of entropy or global minimum of energy);
(ii) they can relax towards a metastable state (local maximum of
entropy or local minimum of energy); (iii) they do not relax towards a
steady state and develop structures at smaller and smaller scales. In
the present situation, we have seen that there are no stable and
metastable states. Therefore, the stability analysis of
Sec. \ref{Analytics} predicts that the system should cascade towards
smaller and smaller scales without limit (except the one fixed by the
finite resolution of the simulations). This is a possible regime (see top of Fig. \ref{fig:ex_relaxinst}) but this is not what is generically
observed in the experiments where long-lived structures at large
scales are found (like at the bottom of Fig. \ref{fig:ex_relaxinst}). Here, we explore
the possibility that these long-lived structures are saddle points of
entropy or energy with relevant constraints. These saddle points are
steady states of the relaxation equations. Although they are unstable
(strictly speaking), we argue that these saddle points can be
long-lived and relatively robust (this idea was previously developed for 2D flows in \cite{naso09a}). Indeed, they are unstable only for certain
(dangerous) perturbations, but not for all perturbations. Therefore,
they can be stable as long as the system does not explore
dangerous perturbations that destabilize them. Of course, the rigorous
characterization of this form of stability is extremely complex. In
order to test this idea in a simple manner, we shall use the
relaxation equations and study the robustness of the saddle points
with respect to them.

\subsection{Numerical method}
\label{numet}

\subsubsection{Generalities}

Our stability analysis is based on the numerical integration of the relaxation equations
\begin{eqnarray}
&\frac{\partial \xi}{\partial t} =  -\chi (\beta \psi + \mu \sigma ),\label{rel1}\\
&\frac{\partial \sigma}{\partial t}  =  -D \left (\beta \displaystyle\frac{\sigma}{r^2} + \mu \xi + \alpha \right ),
\label{eq:canorelax}
\end{eqnarray}
where $D$ and $\chi$ are given functions of $r$ and $z$, and $\beta$, $\mu$ and $\alpha$ evolve in time (see below) so as to guarantee the conservation of the invariants.

In the canonical ensemble, the temperature $\beta$ is fixed and the conserved quantities are the helicity and the angular momentum. The Lagrange multipliers  $\mu(t)$ and $\alpha(t)$ are computed at each time so as to guarantee the conservation of $H$ and $I$. One may check that they are solutions of the system of algebraic equations (see Paper I)
\begin{eqnarray}
&\langle D\xi\rangle \alpha (t) + \left( \langle\chi\sigma^2\rangle +\langle D\xi^2\rangle \right) \mu (t) = -\beta \left( \langle\chi \psi\sigma\rangle  +\langle D\xi\displaystyle\frac{\sigma}{r^2}\rangle \right) \label{eq:coefrelaxcano1} \\
&\langle D \rangle \alpha (t) + \langle D \xi\rangle \mu (t) = -\beta \langle\displaystyle\frac{D\sigma}{r^2}\rangle \label{eq:coefrelaxcano2} .
\end{eqnarray}
 These relaxation equations are associated with the maximization problem (\ref{bes2}) provided that, at any given time, the distribution of angular momentum is given by Eq. (\ref{vp1}) with constant $\beta$ (see Paper I for details). By properly redefining the Lagrange multipliers, they are also associated with the minimization problem (\ref{res13b}).

In the microcanonical ensemble, the conserved quantities are $E$, $H$ and $I$. In the sequel, it will be convenient to fix the time dependence of $\beta$ by imposing
$\beta^{-1}(t)=2\left(E - E^{c.g}(t)\right) $ at each time. Taking into account the two other invariants, one may check that
 $\beta(t)$, $\alpha(t)$ and  $\mu (t)$ are solution of the system of algebraic equations
\begin{eqnarray}
\langle D\xi\rangle \alpha (t) + \left( \langle\chi\sigma^2\rangle + \langle D\xi^2\rangle \right) \mu (t) + \left( \langle\chi\psi\sigma\rangle  +\langle D\displaystyle\frac{\xi\sigma}{r^2}\rangle \right) \beta (t) & = & 0, \label{eq:coefrelaxmicro1}\\
\langle D \rangle \alpha (t) + \langle D \xi\rangle \mu (t) + \langle\displaystyle\frac{D \sigma}{r^2}\rangle \beta (t) & = & 0, \label{eq:coefrelaxmicro2}\\
2 \left(E -E^{c.g}\right) \beta (t) & = & 1. \label{eq:coefrelaxmicro3}
\end{eqnarray}
These relaxation equations are associated with the maximization problem (\ref{bes1}) provided that, at any given time, the distribution of angular momentum is given by Eq. (\ref{vp1}) with $\beta=\beta(t)$ (see Paper I for details).

\begin{figure}[!htb]
\begin{minipage}{0.99\columnwidth}
\includegraphics[width=0.15\columnwidth]{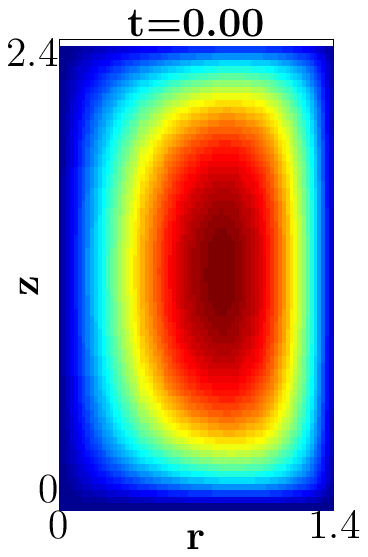}
\includegraphics[width=0.15\columnwidth]{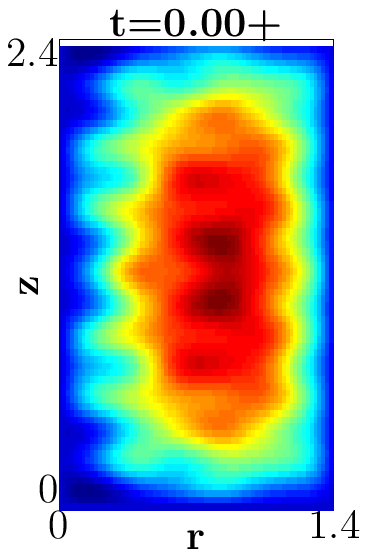}
\includegraphics[width=0.15\columnwidth]{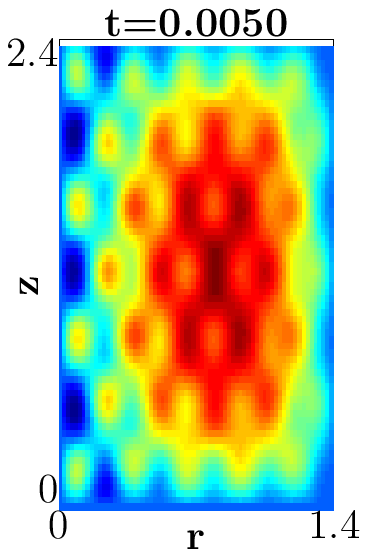}
\includegraphics[width=0.15\columnwidth]{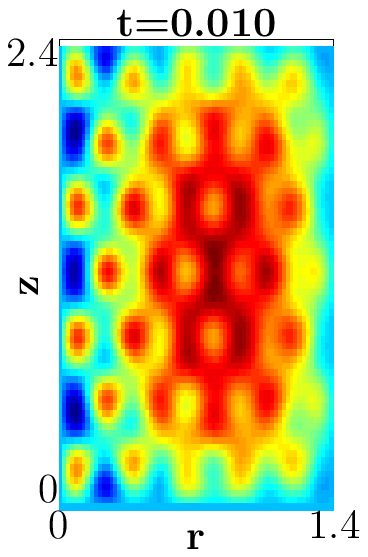}
\includegraphics[width=0.15\columnwidth]{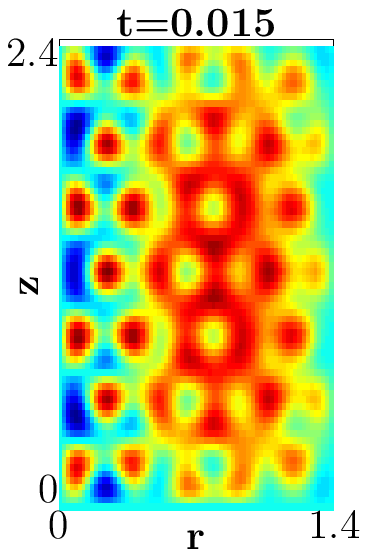}
\includegraphics[width=0.15\columnwidth]{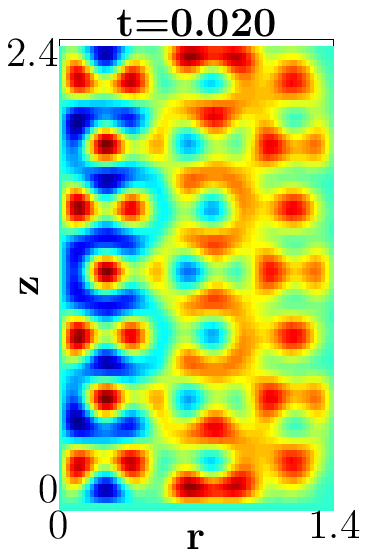}
\end{minipage}
\begin{minipage}{0.99\columnwidth}
\includegraphics[width=0.15\columnwidth]{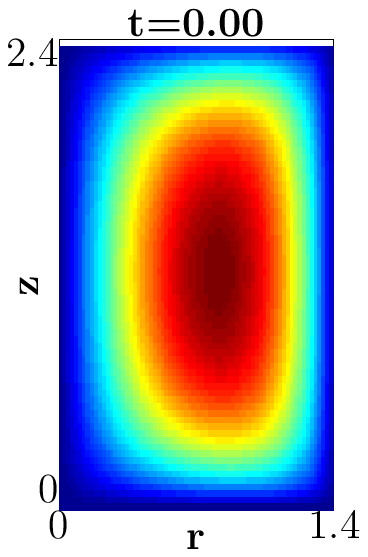}
\includegraphics[width=0.15\columnwidth]{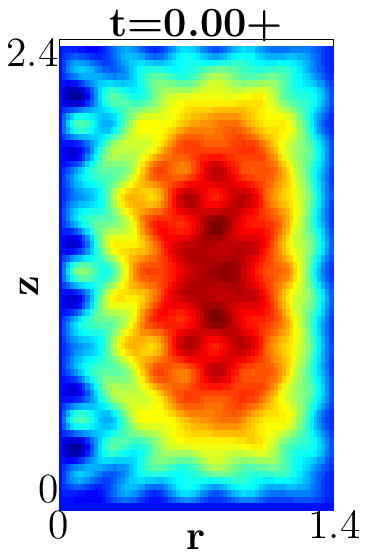}
\includegraphics[width=0.15\columnwidth]{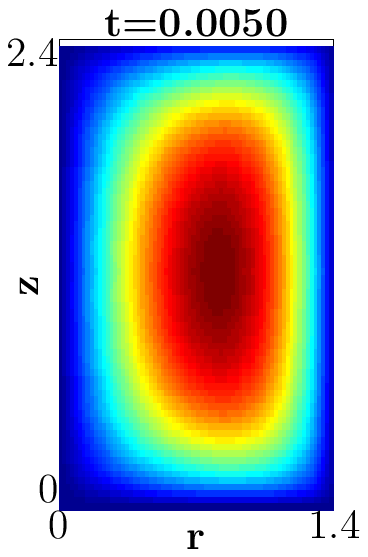}
\includegraphics[width=0.15\columnwidth]{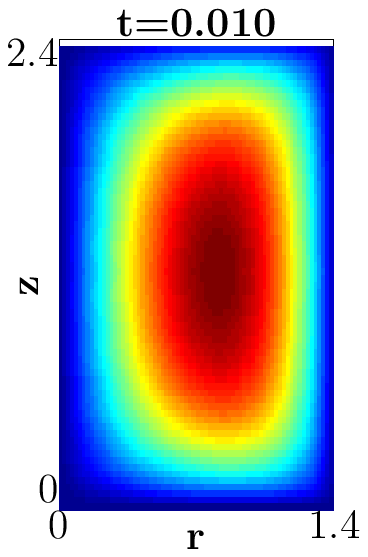}
\includegraphics[width=0.15\columnwidth]{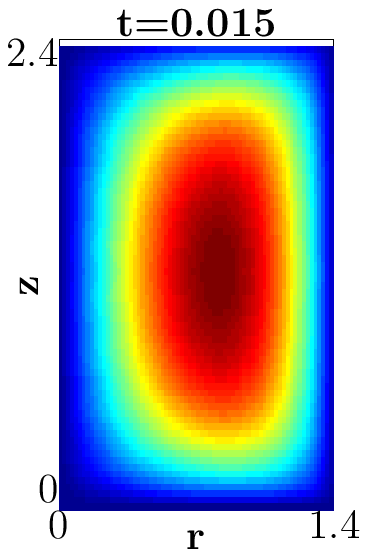}
\includegraphics[width=0.15\columnwidth]{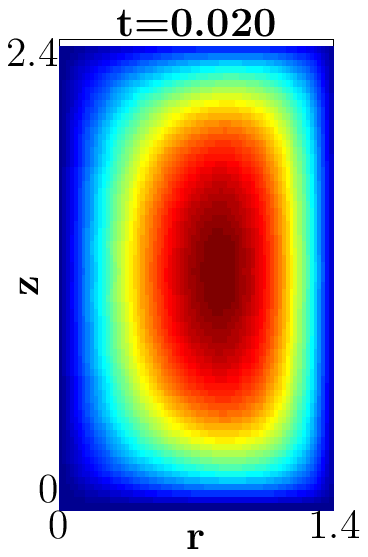}
\end{minipage}
\caption{Time evolution of the stream function of a solution of branch 1 of the continuum. Top: in the fragile case;  Bottom:  in the robust case.}
\label{fig:ex_relaxinst}
\end{figure}

\begin{figure}[!htb]
    \centering
        \includegraphics[width=0.35\columnwidth]{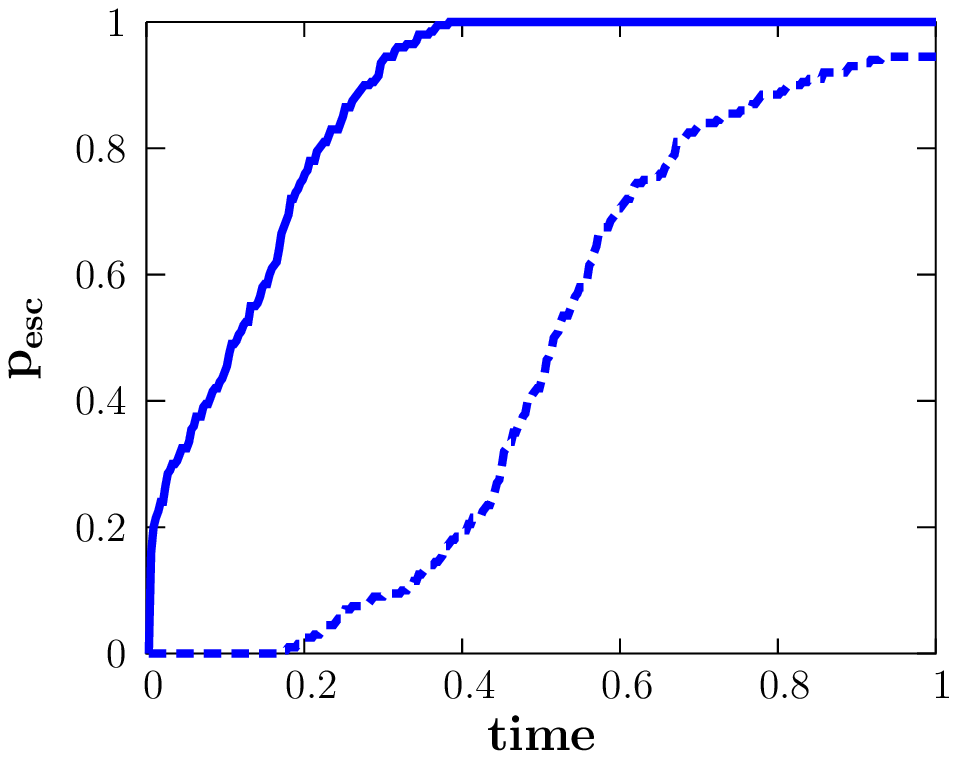}
		\includegraphics[width=0.35\columnwidth]{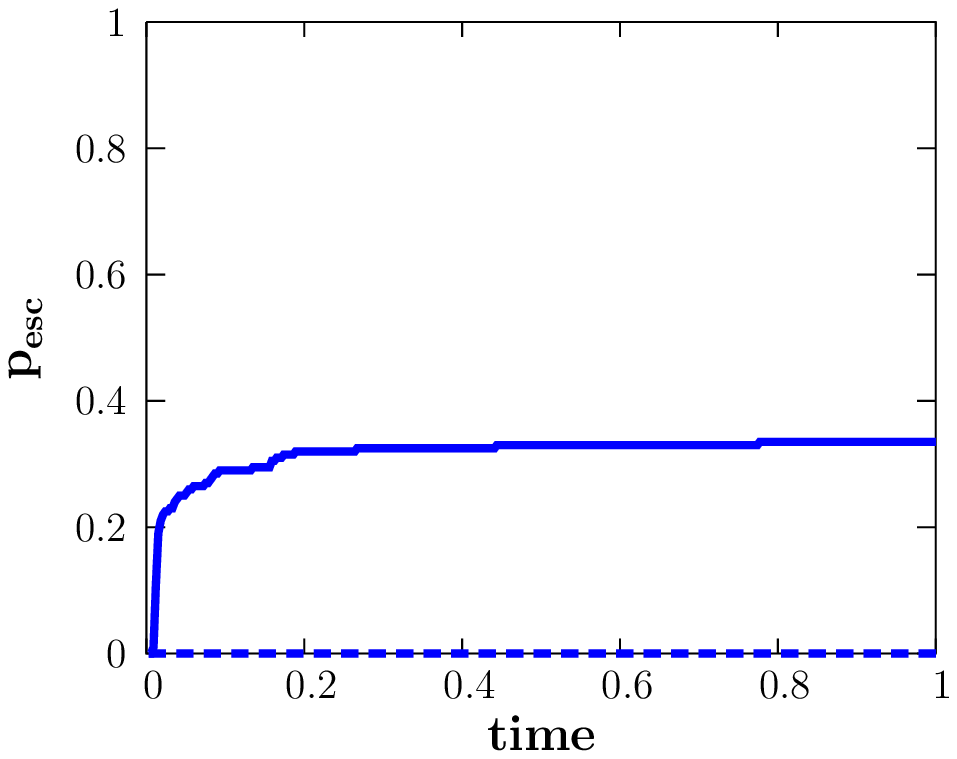}

\caption{Time evolution of $p_{esc}$ made with 200 perturbations around two Beltrami states. Top: fragile solution; Bottom: robust solution. Two different thresholds are used: $\delta = 1.1$ (continuous line)   $\delta = 20$ (dotted line).}
\label{fig:criteres}
\end{figure}

In the sequel we focus on the special case $D=D_* r^2$ and $\chi=\chi_* r^{-2}$, where $D_*$ and $\chi_*$ are constants, that allows a simple numerical treatement of the relaxation equations by projection along the Beltrami eigenmodes: 
\begin{eqnarray}
&&\sigma = \sum_1^{N_m} r s_{\bf n}\phi_{\bf n},\label{dec1}\\
&&\xi = \sum_1^{N_m} r^{-1} x_{\bf n}\phi_{\bf n},\label{dec2}\\
&&\psi = \sum_1^{N_m} r p_{\bf n}\phi_{\bf n},
\label{eq:decomp_std}
\end{eqnarray}
where ${\bf n}=(m,n)$, $0\le n\le N$, $0\le m\le M$ label the modes and $N_m=N\times M$ is the number of modes.
In that case,  Eqs. (\ref{rel1}) and (\ref{eq:canorelax}) can be transformed into a set of $N_m$ ODEs: 
\begin{eqnarray}
&&\dot{s_{\bf n}} = -D_* \left[ \beta s_{\bf n} +\mu x_{\bf n} +\alpha \langle r\phi_{\bf n}\rangle \right], \label{ode1} \\
&&\dot{x_{\bf n}} = -\chi_* \left[ \beta p_{\bf n} +\mu s_{\bf n}\right],\label{ode2}\\
&&p_{\bf n} = B_{\bf n}^{-2} x_{\bf n},
\label{eq:relax_simp_proj}
\end{eqnarray}
where $B_{\bf n}$ is such that $-\Delta \phi_{\bf n}+\phi_{\bf n}/r^2=B_{\bf n}^{2}\phi_{\bf n}$. Note that the constraints couple Eqs. (\ref{ode1})-(\ref{eq:relax_simp_proj}) through the parameters $\beta$, $\alpha$ and $\mu$.
To investigate the robustness of a given stationary solution, we first
perturb it with a suitable perturbation (see below), and then follow
its dynamics thanks to the relaxation equations. Two typical time
evolutions are provided in Fig \ref{fig:ex_relaxinst}: if the solution
is fragile with respect to the perturbation, it will cascade to
another solution (usually the solution of smallest scale permitted by
our resolution); if the solution is robust with respect to this
perturbation, it will eventually return to its initial unperturbed
state.
To quantify the robustness of a given solution, we define a probabilistic stability criterion by computing the probability for the solution ``to escape'' from its basin of attraction. To that purpose, we select a threshold $\delta$ and compute at each time the probability of escape
\begin{equation}
p_{esc}(t)  = {\rm Prob} \left[ \displaystyle\frac{\langle r^{-2}(\delta \sigma (t))^2 \rangle  }{\langle r^{-2}(\delta \sigma (0^{+}))^2\rangle } > \delta \right],
\end{equation}
using $N_p$ realizations with perturbations drawn at random at $t=0$  from a suitable ensemble (see below). This allows us to define ``statistically fragile'' solutions as those for which
$p_{esc}(t)\to 1$ when $t\to\infty$, the others being referred to as ``statistically robust''. In practice, the limit $t\to\infty$ is not accessible. We thus generalize this notion to a "finite time", by considering the asymptotic value of $p_{esc}$ reached at the largest time of the simulation, $t_{max}$.
 In addition, the asymptotic value of $p_{esc}(t)$ provides a mean to quantify the degree of robustness of a solution. Examples are given in Fig. \ref{fig:criteres}, for a fragile and for a robust solution. As can be seen, the fragile solution is fragile whatever the threshold $\delta$. However, the degree of robustness of a solution depends on the threshold $\delta$. Quite naturally, the larger the threshold, the more robust the solution.

{\it Remark}: although the variational problems (\ref{bes1}), (\ref{bes2}) and  (\ref{res13b}) are equivalent, and all lead to the absence of stable equilibrium state, the corresponding relaxation equations described previously are different. Therefore, the robustness of the saddle points will be different in the canonical and microcanonical settings. This can be viewed as a form of ``ensembles inequivalence'' for an out-of-equilibrium situation.

\subsubsection{Perturbations}

The stability must be investigated using perturbations that rigorously conserve the integral constraints. This puts some conditions regarding the shape of the possible perturbations that we can use.

In the canonical ensemble, the integral constraints are $H$ and $I$. Given an initial stationary solution $(\sigma^\star, \xi^\star)$, the perturbations $(\delta \sigma,\delta \xi)$ must obey
\begin{eqnarray}
&& \langle\xi^\star\delta \sigma\rangle  + \langle\delta \xi\sigma^\star\rangle  + \langle\delta \xi\delta \sigma\rangle  =0, \\
&& \langle\delta \sigma\rangle  = 0.
\label{eq:contraintescano}
\end{eqnarray}
One can check that this set of constraints is satisfied by any perturbation of the form 
\begin{eqnarray}
&&\delta \sigma = {\epsilon} r  \left[ \phi_{\bf i_1} - \displaystyle\frac{\langle r\phi_{\bf i_1}\rangle }{\langle r\phi_{\bf i_0}\rangle } \phi_{\bf i_0}\right]\\
&&\delta \xi = {\epsilon} r^{-1} {s_{\bf i_2}^\star}^{-1}\left[\displaystyle\frac{\langle r\phi_{\bf i_1}\rangle }{\langle r\phi_{\bf i_0}\rangle }x^\star_{\bf i_0}-x^\star_{\bf i_1}\right]\phi_{\bf i_2}\\
&&\delta \psi = {\epsilon} r  {s_{\bf i_2}^\star}^{-1} B_{\bf i_2}^{-2}\left[\displaystyle\frac{\langle r\phi_{\bf i_1}\rangle }{\langle r\phi_{\bf i_0}\rangle }x^\star_{\bf i_0}-x^\star_{\bf i_1}\right]\phi_{\bf i_2}
\label{eq:perturb}
\end{eqnarray}
 where ${\epsilon}$ is the amplitude of the perturbation, ${\bf i_0}$ labels an even mode while  ${\bf i_1}$ and ${\bf i_2}$  label two different modes  different from ${\bf i_0}$ such that  $\langle  \sigma^\star \phi_{\bf i_2}\rangle  \neq 0 $. Following Eqs. (\ref{dec1}) and (\ref{dec2}), we have set $ \sigma^\star =  \sum r s^\star_{\bf i} \phi_{\bf i}$ and $\xi^\star = \sum r^{-1}x^\star_{\bf i}\phi_{\bf i}$.
In the sequel, we fix the amplitude of the perturbation ${\epsilon}$ through the norm
 $A_p^2 = \langle  r^{-2}(\delta \sigma)^2 \rangle  $ by imposing
$${\epsilon} = A_p \left[1+ \left(\displaystyle\frac{\langle r\phi_{\bf i_1}\rangle }{\langle r\phi_{\bf i_0}\rangle }\right)^2\right]^{-\frac{1}{2}}.$$
The modes ${\bf i_0}$, ${\bf i_1}$ and ${\bf i_2}$ are chosen randomly according to the following procedure: i) we draw ${\bf i_0}$ following a uniform law among the $N\times M$ even modes; ii) we draw ${\bf i_2}$ following a uniform law among the $N\times M$ or $N\times M+1$ modes of the set of allowed $\sigma^\star$, excluding ${\bf i_0}$. This mode is therefore necessarily even for solution of continuum, and often even for mixed solutions; iii) we draw ${\bf i_1}$ following a uniform law among the $2\times N\times M$ even and odd modes, excluding ${\bf i_0}$ and ${\bf i_2}$.
This choice allows the generation of  $N_p$ random perturbations with the  same amplitude $A_p$.

In the microcanonical ensemble, the relaxation equations conserve in addition the energy. To satisfy this additional constraint, we choose the perturbations according to the same procedure as in the canonical case, and then determine the initial value of the temperature $\beta_0=\beta(t=0)$ in order to guarantee the conservation of the energy \footnote{This procedure is permissible  because the helicity $H$ and angular momentum $I$ determine the mean flow while the energy $E$ determines the temperature.}. As explained previously, this amounts to taking
$$\beta_0^{-1}=2\left(E - E^{c.g}(\sigma^\star + \delta \sigma,\xi^\star + \delta \xi)\right). $$
In the following, we shall group the perturbations into subclasses such that perturbations of the same class have the same temperature $\beta_0$ or, equivalently, the same macroscopic energy $E^{c.g}(\sigma^\star + \delta \sigma,\xi^\star + \delta \xi)$. Note that the initial temperature of the perturbation differs from the temperature of the equilibrium state which is given by
$$\beta_{eq}^{-1}=2\left(E - E^{c.g}(\sigma^\star,\xi^\star)\right). $$

\subsubsection{Parameters}

In the sequel, we focus on the stability analysis in the case L, for the first three branches of solutions, relevant for comparison with experiments, see Paper III \cite{dubrulle09}. Our parameters are as follows:
\begin{itemize}
\item The number of modes is $N_m = 2\ N M$ with $N = 10$ (radial modes) and $M = 12$ (vertical modes) corresponding to 120 even modes and 120 odd modes. The radial and vertical lengths are $R = 1.2$ and $h = 1.4$. 
\item The amplitude of the perturbations is  $A_p = 0.05$. We consider  $N_p = 200$ realizations  for each given stationary solution.
\item The parameters $D_*$ and $\chi_*$ are both taken equal to $1$. The relaxation equations are integrated using an implicit Heun scheme. The time step is empirically chosen proportional to  $\beta(t=0)^{-1}$. For $\beta(t=0) = 1$, the time step  is 0.02. We have checked that this time step is small enough to guarantee
the numerical conservation of $I$ and $H$ (canonical case) or $E$, $I$ and $H$ (microcanonical case).
\end{itemize}

\subsubsection{Numerical protocol}

For any value of   $\Lambda$ on a given branch of solutions, we proceed as follows:
\subparagraph*{Canonical ensemble}
\begin{itemize}
\item[i)] We fix the value of the temperature $\beta$ (it remains constant during the evolution). In the sequel, we focus on five arbitrary values, $\beta =1000,100,10,1,0.1$, chosen so as to span a wide range.
\item[ii)] We compute the Beltrami solution  $(\sigma^\star,\xi^\star)$ corresponding to a prescribed value of $\Lambda$ on the given branch.
\item[iii)] We generate  $N_p$  perturbed initial conditions leaving unchanged the  helicity and the angular momentum of $(\sigma^\star,\xi^\star)$.
\item[iv)] We evolve the perturbed initial conditions through Eqs. (\ref{ode1})-(\ref{eq:relax_simp_proj}) and  Eq. (\ref{eq:coefrelaxcano1},\ref{eq:coefrelaxcano2}) for a certain amount of time $t_{max}$.
\end{itemize}

\subparagraph*{Microcanonical ensemble}
\begin{itemize}
\item[i)] We fix the value of the energy $E$ (it remains constant during the evolution). In the sequel, it is fixed after arbitrary choice of five  values of the temperature, $\beta_0 =1000,100,10,1,0.1$, chosen so as to span a wide range. Once $\beta_0$ has been fixed, the total energy is then fixed. It can vary from one realization to the other, but does not vary along the evolution.
\item[ii)] We compute the stationary  Beltrami solution $(\sigma^\star,\xi^\star)$ corresponding to a prescribed value of $\Lambda$ on the given branch.
\item[iii)] We generate  $N_p$  perturbed initial conditions leaving unchanged the  helicity and the angular momentum of $(\sigma^\star,\xi^\star)$.
\item[iv)] We group together the perturbations that have the same initial temperature $\beta_0^{-1}=2(E-E^{c.g.})$ measuring the initial energy of the fluctuations (equivalently, these perturbations have the same value of macroscopic energy $E^{c.g.}$).
\item[iv)]  We evolve the perturbed initial condition through Eqs. (\ref{ode1})-(\ref{eq:relax_simp_proj}) and (\ref{eq:coefrelaxmicro1}-\ref{eq:coefrelaxmicro3}) for a certain amount of time $t_{max}$.
\end{itemize}

\subsection{Numerical results}
\label{Numerics}

\subsubsection{Robustness in the canonical ensemble}

\begin{figure}[!htb]
\begin{minipage}{0.99\columnwidth}
\includegraphics[width=0.32\columnwidth]{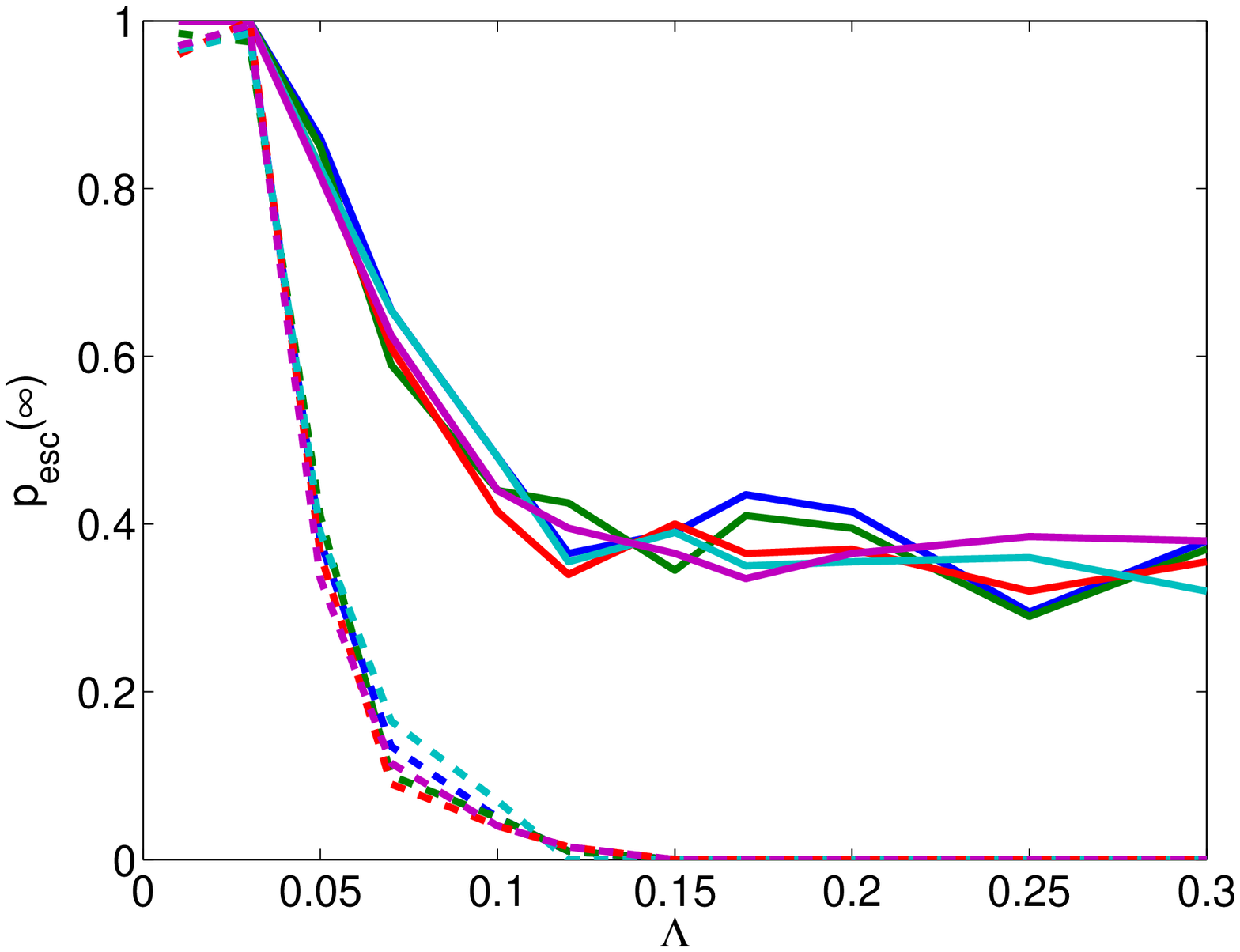}\label{fig:micro_branche1}
\includegraphics[width=0.32\columnwidth]{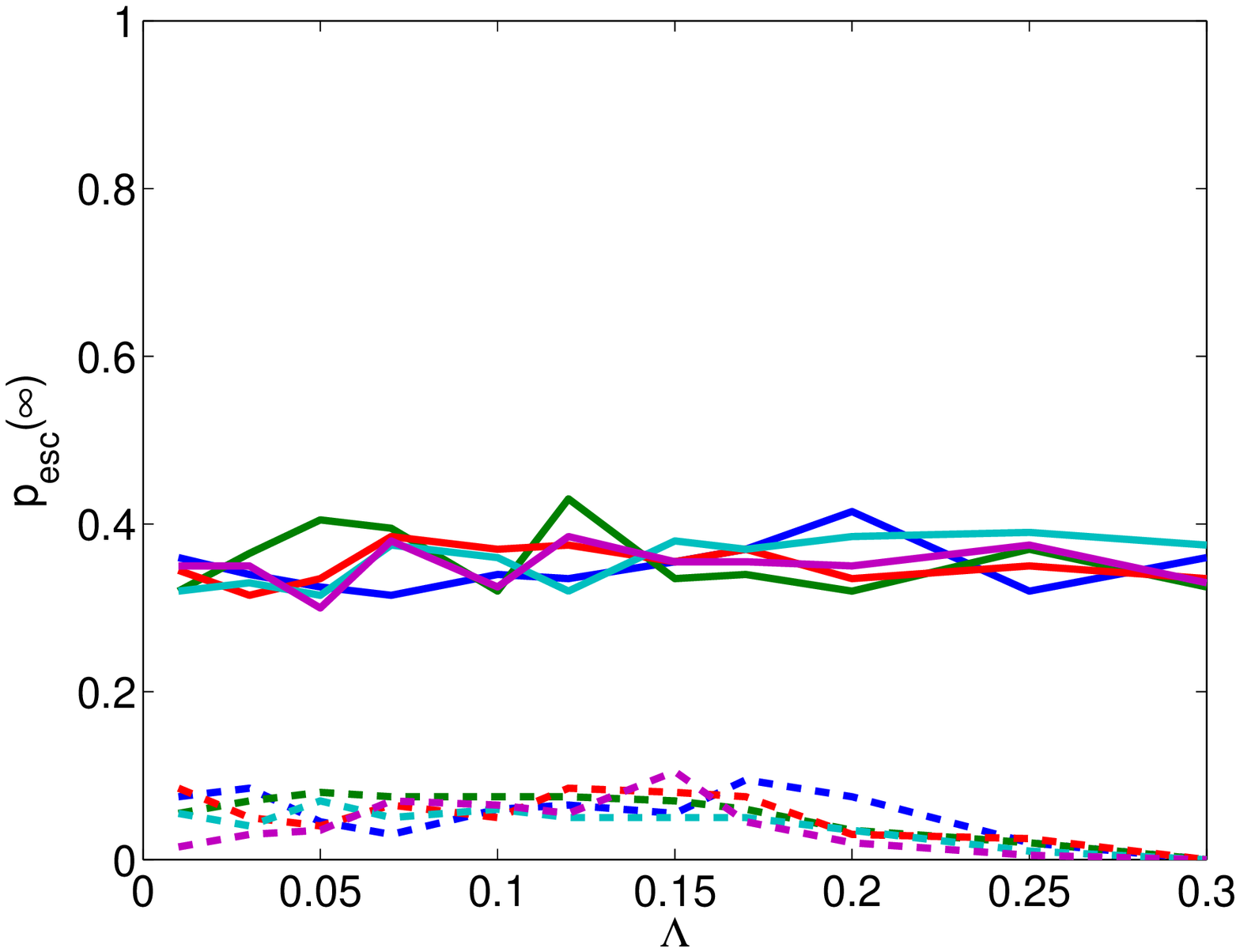}\label{fig:micro_branche2}
\includegraphics[width=0.32\columnwidth]{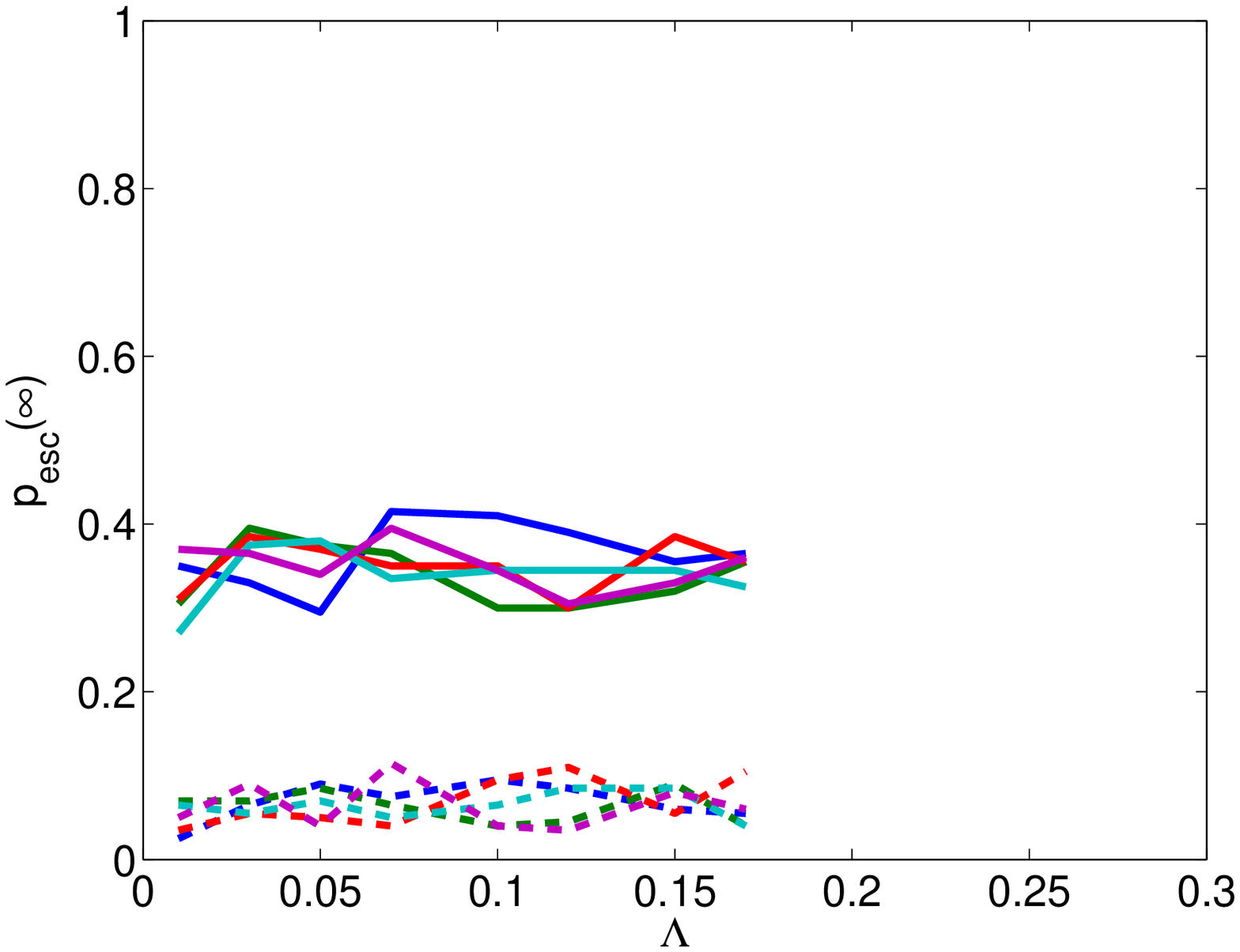}\label{fig:micro_mixte}
\end{minipage}
\caption{\textbf{Canonical ensemble}. $p_{esc}(\infty)$ as function of $\Lambda$ on the three branches
(left: branch 1, middle: branch 2, right: mixed branch) at different temperatures $1/\beta$:  $0.001 (\textcolor{blue}{\bullet})$, $0.01 (\textcolor{green}{\bullet})$, $0.1 (\textcolor{red}{\bullet})$, $1 (\textcolor{cyan}{\bullet})$, and $10 (\textcolor{magenta}{\bullet})$. Two different thresholds are used: $\delta = 1.1$ (continuous line)   $\delta = 20$ (dotted line).}
\label{fig:inequivalence_cano}
\end{figure}

\begin{figure}[!htb]
    \centering
        \includegraphics[scale=0.5,width=0.45\columnwidth]{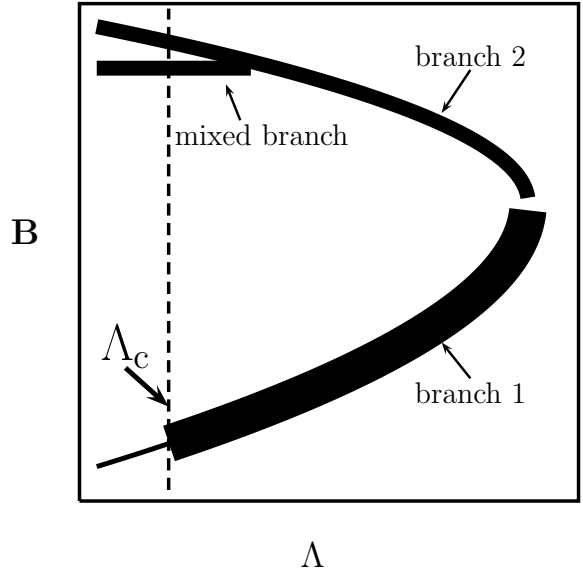}
		
\caption{Robustness of the three branches in the canonical case computed at a given $t_{max}$. The lines are increasingly fat with increasing $p_{esc}(\infty)$, i.e. robustness. Note that the value of $\Lambda_c$ increases with increasing $t_{max}$.}
\label{fig:stabcano}
\end{figure}

On the three branches, we computed the value $p_{esc}(\infty)$ (computed at $t_{max}$) as a function of $\Lambda$ for different temperatures $T=\beta^{-1}$. The results are displayed on Fig. \ref{fig:inequivalence_cano}.
The mixed branch (vertical dipoles) and branch 2 (reversed monopoles) are found very robust for high threshold, and still retain a certain degree of robustness for a small threshold, with a 40 per cent probability of escape.   There is no clear dependence on the temperature. This is natural, since temperature can be eliminated by a suitable rescaling of time  (or of coefficients $D$ and $\chi$) and redefinition of Lagrange parameters.
The behavior on branch 1 (direct monopoles) is more contrasted and provides a very clear transition around the critical value $\Lambda_c\approx 0.1$. For $\Lambda <  \Lambda_c$, the probability to escape is close to 1, meaning large fragility of the branch. For $\Lambda  > \Lambda_c$, the branch becomes much more robust, reaching a larger degree of robustness than the two other branches for high threshold, while reaching the same robustness for small threshold. The different behaviors are summarized on Fig. \ref{fig:stabcano}.

Note that the value $\Lambda_c=0.1$ is somewhat arbitrary. Indeed, increasing $t_{max}$ further, we observed the same qualitative scenario, with an increased value of $\Lambda_c$.
We also observed that over sufficiently long time, the branch 2 tends to become unstable, past a value of the order $\Lambda=0.25$.

\subsubsection{Robustness in the microcanonical ensemble}

\begin{figure}[!htb]
\begin{minipage}{0.99\columnwidth}
\includegraphics[width=0.32\columnwidth]{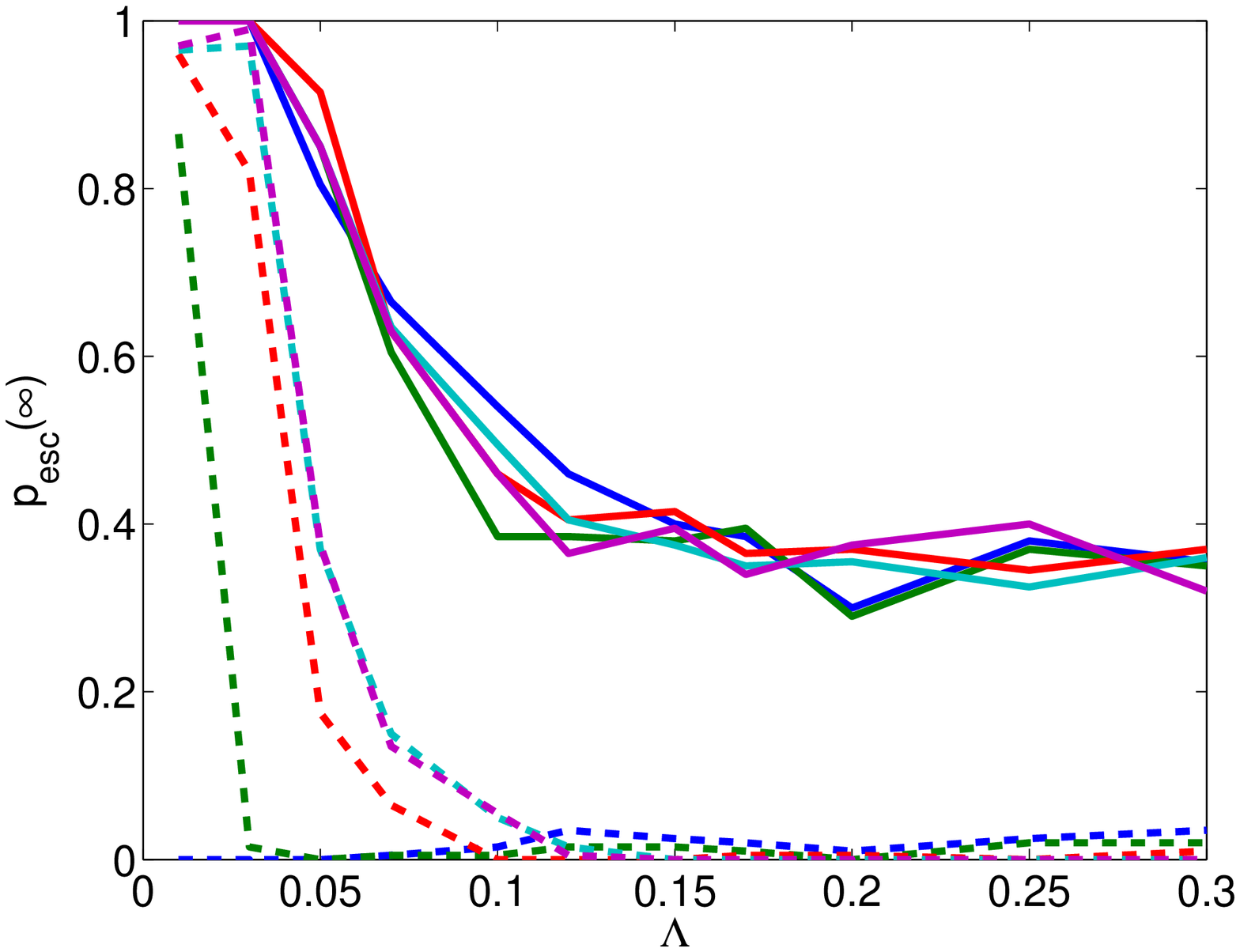}
\includegraphics[width=0.32\columnwidth]{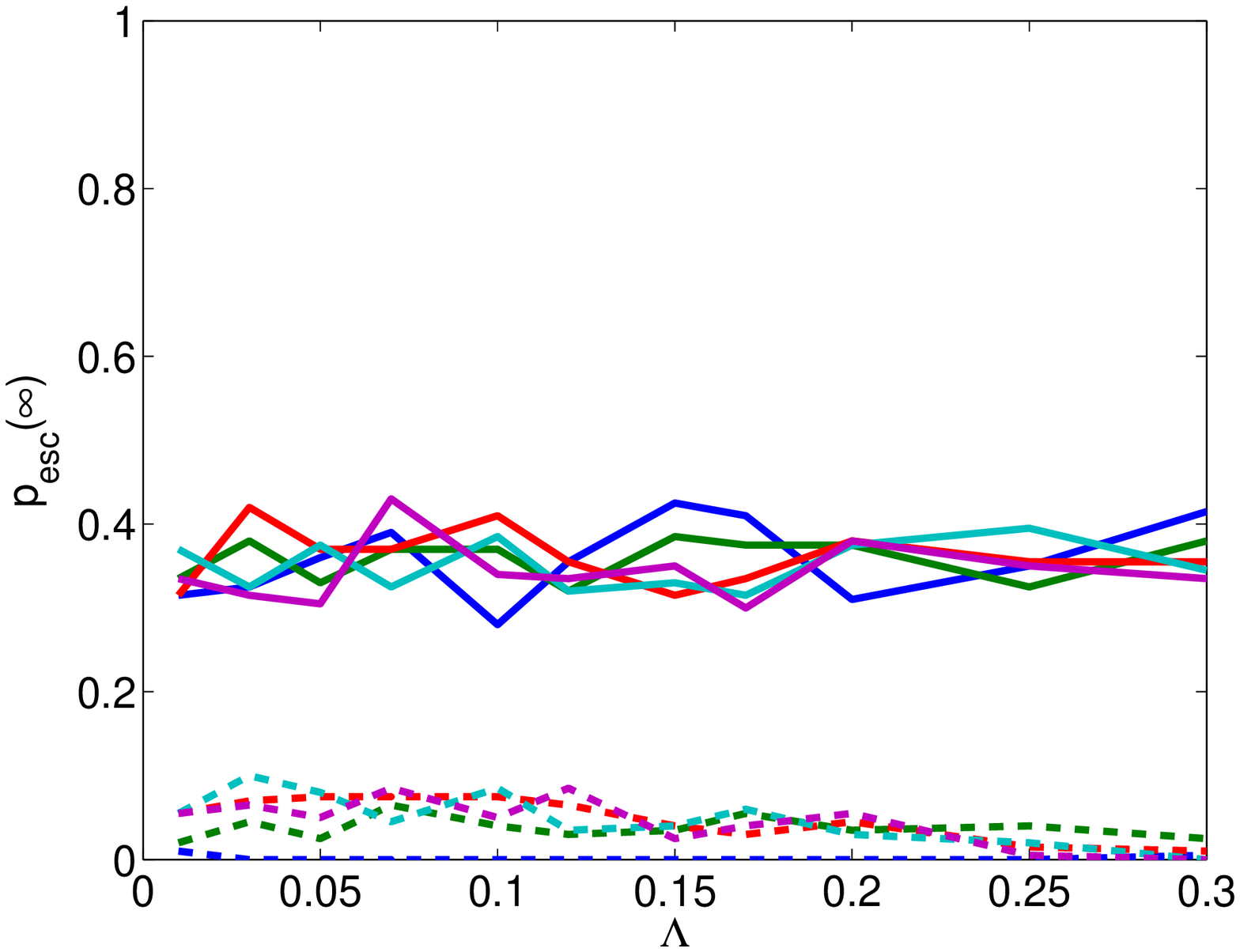}
\includegraphics[width=0.32\columnwidth]{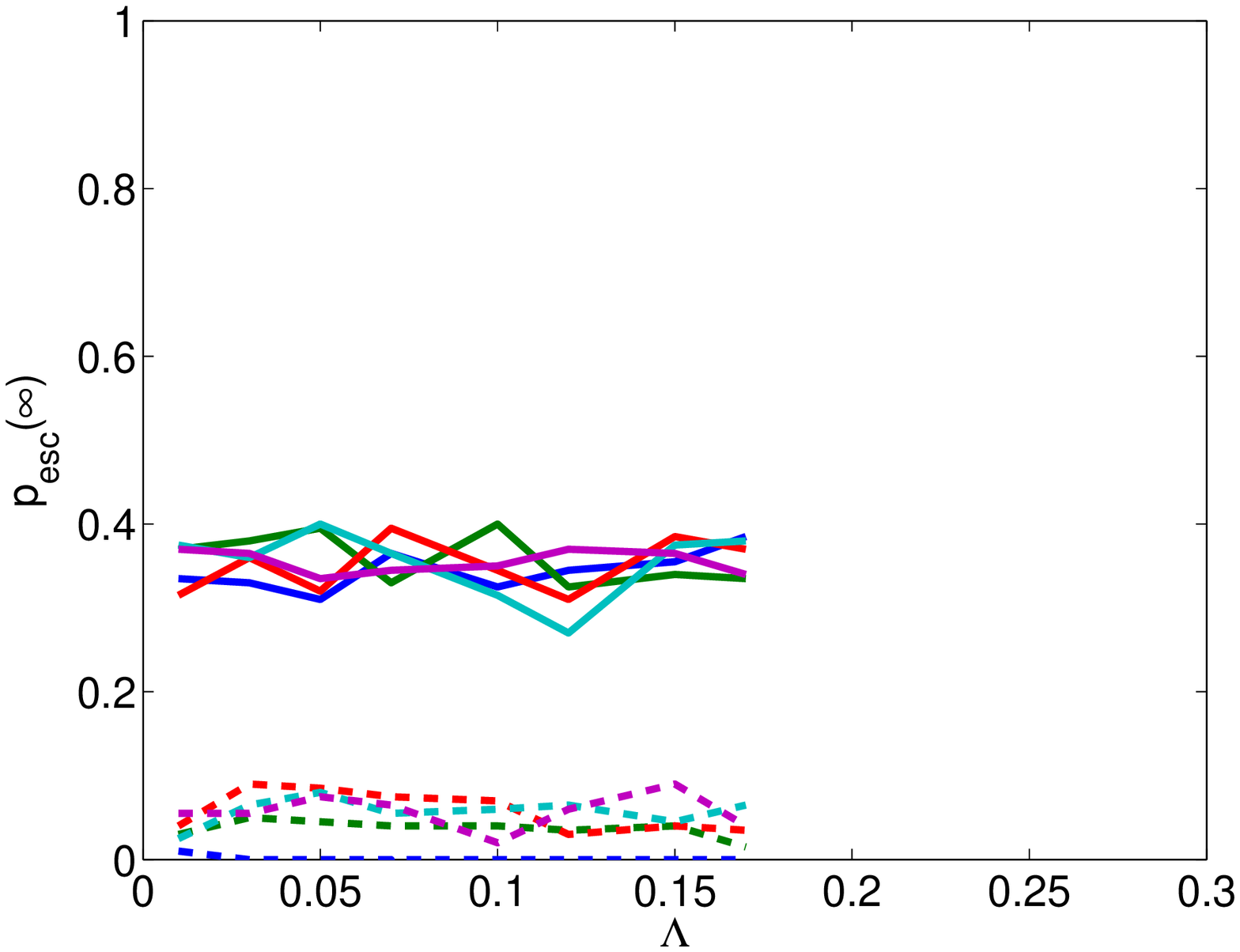}
\end{minipage}
\caption{\textbf{Microcanonical ensemble}. $p_{esc}(\infty)$ as function of $\Lambda$ on the three branches (left: branch 1; middle: branch 2; right: mixed branch) at different initial temperatures $1/\beta_0$:  $0.001 (\textcolor{blue}{\bullet})$, $0.01 (\textcolor{green}{\bullet})$, $0.1 (\textcolor{red}{\bullet})$, $1 (\textcolor{cyan}{\bullet})$, and $10 (\textcolor{magenta}{\bullet})$. Two different thresholds are used: $\delta = 1.1$ (continuous line)   $\delta = 20$ (dotted line).}
\label{fig:inequivalence_micro}
\end{figure}

\begin{figure}[!htb]
\begin{minipage}{0.99\columnwidth}
\includegraphics[width=0.45\columnwidth]{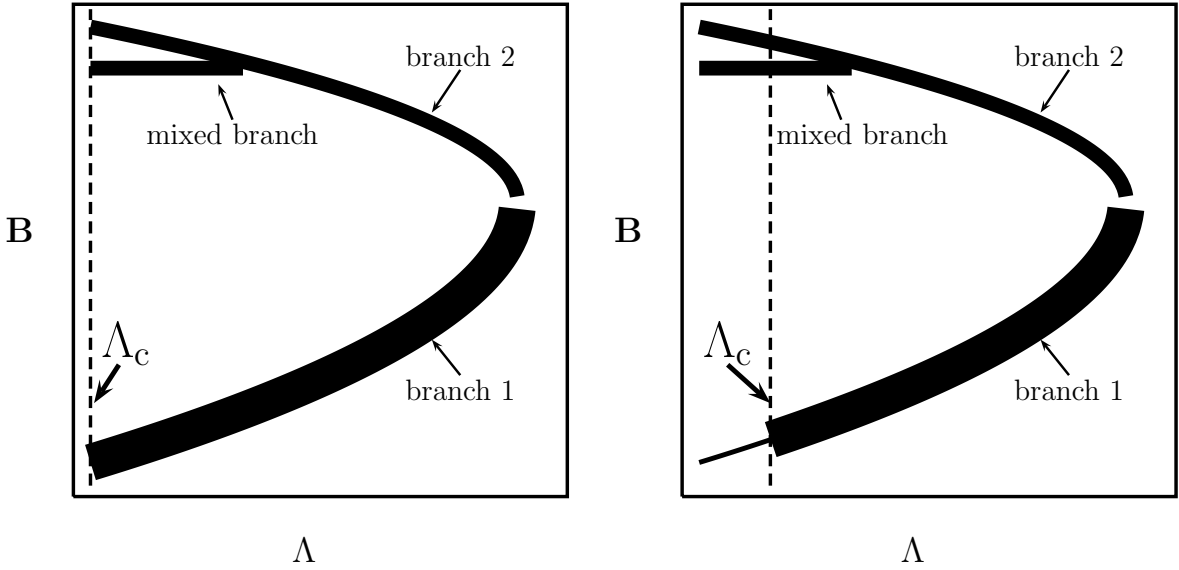}\label{fig:stab0}
\includegraphics[width=0.45\columnwidth]{lam1cor.eps}\label{fig:stab1}
\end{minipage}
\caption{Robustness of the three branches in the microcanonical case 
computed at a given $t_{max}$. Left: for a low initial temperature; right: for a high initial temperature. The lines are increasingly fat with increasing $p_{esc}(\infty)$, i.e. robustness. Note that the value of $\Lambda_c$ increases with increasing $t_{max}$.}
\label{fig:stabmicro}
\end{figure}

On the three branches, we computed the value $p_{esc}(\infty)$ (computed at $t_{max}$) as a function of $\Lambda$ for classes of perturbations with different initial temperature $T_0=\beta_0^{-1}$. Note that the initial temperature fixes the amplitude of the velocity fluctuations. The results are displayed on Fig. \ref{fig:inequivalence_micro}. For the mixed branch (vertical dipoles) and branch 2 (reversed monopoles), the microcanonical results do not noticeably differ from the canonical results: the two branches are found very robust for large threshold, and still retain a certain degree of robustness for a small threshold, with a 40 per cent probability of escape.  There is no clear dependence on the initial temperature. There is therefore no ensembles inequivalence for these two branches.
This is not true anymore for branch 1 (direct monopoles). Indeed, one still observes a transition from robustness to fragility around a critical value $\Lambda_c$ but this quantity depends on the initial temperature $T_0$: it takes a value $\Lambda_c\approx 0.1$ at large initial temperatures (large velocity fluctuations) and then decreases to $0$ for small initial temperatures (small velocity fluctuations). The difference of robustness observed between the two ensembles may be seen as a kind of inequivalence of ensembles at small initial temperatures. The different behaviors are summarized on Fig. \ref{fig:stabmicro}. Like in the canonical case, we checked that an increase of $t_{max}$ results in a larger fragility of the branch 1 and 2 towards small $\Lambda$, at a given temperature.

{\it Remark:} note that perturbations with small initial temperature have large macroscopic energies corresponding to perturbations at large scales. According to Sec. \ref{Analytics} such perturbations are less destabilizing than perturbations at small scales (associated with small macroscopic energies hence large temperatures). This may explain the numerical results.

\section{Discussion}
\label{Discussion}

\subsection{Generalized ensemble inequivalence}

We have studied the thermodynamics of axisymmetric Euler-Beltrami flows and proved the coexistence of several equilibrium states for the same values of the control parameters. All these states are saddle points of entropy but they can have very long lifetime as long as the system does not spontaneously develop dangerous perturbations.  We have numerically explored  the robustness of some of these states by using relaxation equations in the canonical and microcanonical ensembles. The dipoles (mixed branch) and the reversed monopoles (branch 2) were found to be rather robust in both ensembles. Furthermore, in the microcanonical ensemble there is no dependence on the initial temperature on these branches. By contrast, the direct monopoles (branch 1) display a sharp transition around a critical value $\Lambda_c$. The value of $\Lambda_c$ increases with increasing integration time. In the microcanonical ensemble, this value also decreases with decreasing initial temperature, resulting in a difference of robustness in the canonical and microcanonical ensembles. This difference may be seen as a kind of ``ensembles inequivalence''. This is, however, a very unconventional terminology since it concerns here the robustness of {\it saddle points} with respect to random  perturbations that keep the energy or the temperature fixed, over a finite amount of time.

\subsection{Bifurcation scenario}

\begin{figure}[!htb]
\begin{minipage}{0.99\columnwidth}
\includegraphics[width=0.32\columnwidth]{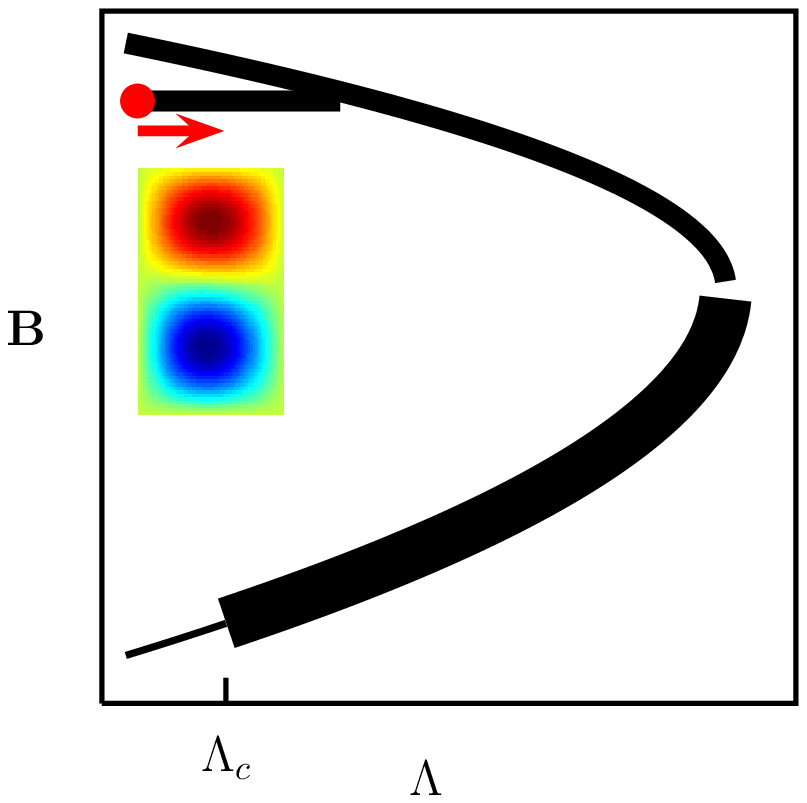}
\includegraphics[width=0.32\columnwidth]{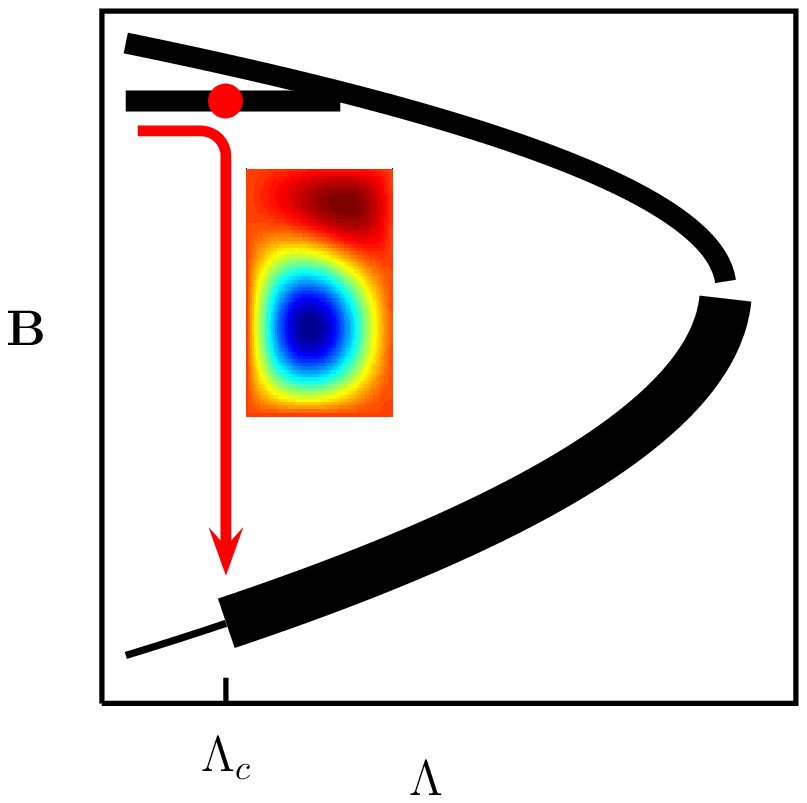}
\includegraphics[width=0.32\columnwidth]{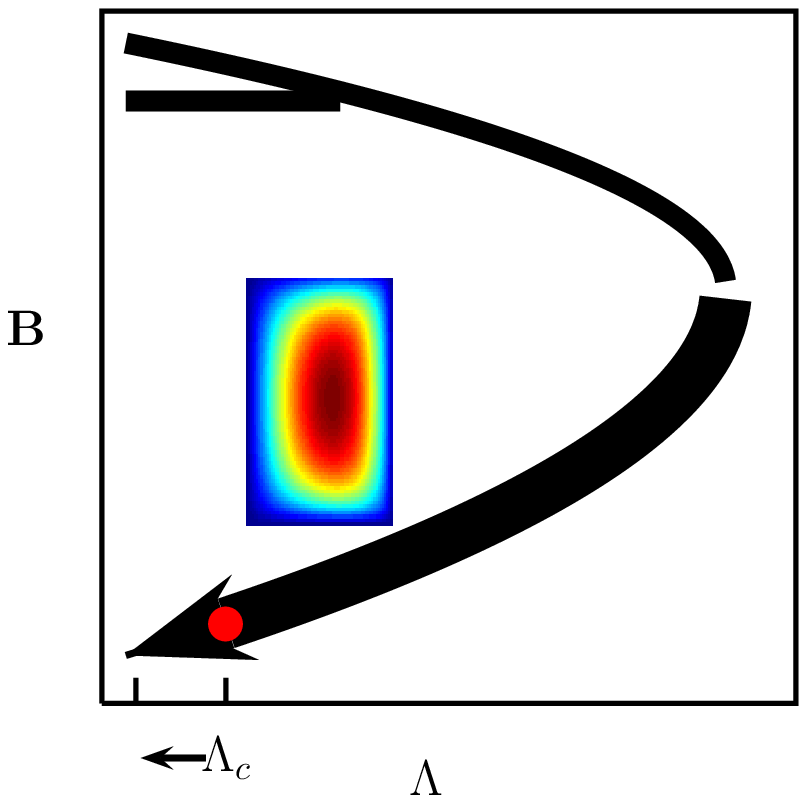}
\end{minipage}
\caption{Bifurcation scenario. Left: the system starts at $\Lambda=0$, on the mixed branch,  in a two-cells topology (vertical dipole); Middle: increasing $\Lambda$ up to $\Lambda_{c}$ (red arrow), the system follows the mixed branch. One cell grows at the expense of the other, resulting in a shift of the mixing layer upwards; Right: At $\Lambda_c$, the system bifurcates towards branch 1 (more stable), resulting in a one-cell topology (direct monopole). In this new state, the fluctuations are much milder (empirical fact from experiments), thereby allowing a stabilization of the branch towards lower $\Lambda$. Therefore, the monopole subsists beyond this point, even after a redecreasing of $\Lambda$, depicted by the black arrow.}
\label{fig:scenarhyst}
\end{figure}

The simulations have shown that the dipole (two-cells solution) is relatively robust for any value of the angular momentum. On the other hand, the direct  monopole  (one-cell solution) is  very fragile at low angular momentum but becomes robust at high angular momentum. In that case, it is even more robust than the dipole. Therefore, increasing the total angular momentum of the flow, one expects to  observe a transition from the two-cells solution (antisymmetric with respect to the middle plane) to the one-cell solution (symmetric with respect to the middle plane). This bifurcation scenario is sketched in Fig. \ref{fig:scenarhyst}. It is reminiscent  of the turbulent transition reported in the von K\'arm\'an flow \cite{ravelet} in which the initial two-cells flow observed at zero global rotation suddenly bifurcates
when the rotation is large enough. Once the bifurcation has taken place, the level of fluctuation is experimentally observed to decrease strongly, resulting in a decrease of the statistical temperature. In our scenario, this means that the monopole branch is suddenly stabilized with respect to redecrease of the total angular momentum of the flow, resulting in a hysteresis that has also been observed experimentally.
It would therefore be  interesting to investigate more closely the relevance of our scenario to the experimental system. This is done in the next paper \cite{dubrulle09}.

\subsection{Richardson energy cascade}

An interesting outcome of our study lies in the fate of the solutions when they are destabilized by a dangerous perturbation: due to the energy minimization principle,
the unstable solution tends to ``cascade'' towards a higher wavenumber solution in a way reminiscent to the Richardson energy cascade of 3D turbulence (see Fig. \ref{fig:ex_relaxinst}). The cascade stops when the largest available wavenumber is reached, since dangerous perturbations are necessarily at smaller scale than the achieved state. This form of energy condensation at the smallest scale may be seen as an interesting counterpart (in the opposite sense) of the large scale energy condensation observed in 2D turbulence via the inverse energy cascade process. This  is the signature of the 2D and a half nature of our system, intermediate between 2D and 3D turbulence.

\section{Conclusion}

We have characterized the thermodynamical equilibrium
states of axisymmetric Euler-Beltrami flows and proved the coexistence of several equilibrium states
 for a given value of the control parameter like
in 2D turbulence \cite{jfm96}. We further showed that all states are saddle points
of entropy and can, in principle, be destabilized by
a perturbation with a larger wavenumber, resulting in a structure at
the smallest available scale. This mechanism is
therefore reminiscent of the 3D Richardson energy cascade
towards smaller and smaller scales. Therefore, our system is truly intermediate between 2D turbulence (coherent structures) and 3D turbulence (energy cascade). Through a numerical exploration of the robustness of the equilibrium
states with respect to random perturbations using a relaxation
algorithm in both canonical and microcanonical
ensembles, we showed however that these saddle points of entropy can
be very robust and therefore play a role in the dynamics.
We evidenced differences in the robustness of the
solutions in the canonical and microcanonical ensembles leading to a theoretical scenario
of bifurcation between two different equilibria (with
one or two cells) that resembles a
recent observation of a turbulent bifurcation in a von K\'arm\'an
experiment \cite{ravelet}.

This work was supported by European Contract WALLTURB.

\appendix

\section*{Appendix A}

We show that the first zero of $F(B)$, denoted $B_*$, is always
between the first $B''_{1}$ and the second $B''_2$ even eigenmode. To that purpose, we note that if $B<B''_1$ then $({B''_{mn}}^2-B^2)> 0$ for any $(m,n)$ so
that $F(B)>0$. There is no discontinuity of $F$ in the interval
$[0,B''_1[$ so that there is no zero in that interval. Consider now the
interval $]B''_1,B''_2[$. In that interval, $F$ is also
continuous and increasing since
\begin{eqnarray}
F'(B)=2B\sum_{mn}\frac{\langle \phi''_{mn}r\rangle^2}{({B''_{mn}}^2-B^2)^2}>0.
\end{eqnarray}
Moreover, for $B\to ({B''_1})^+$, $F(B)\sim \langle\phi''_1
r\rangle^2/ ({B''_1}^2-B^2)\to -\infty$. Similarly, $F(B)\sim
\langle\phi''_2 r\rangle^2/ ({B''_2}^2-B^2)\to +\infty$ when $B\to
({B''_2})^-$. Therefore, there exists a unique value of $B_*$ in the range $]B''_1,B''_2[$, such that $F(B_*)=0$. This shows that the first zero of $F$
lies in between the first two even eigenmodes.  This property remains true for the successive values of $B_*^{(n)}$ and the successive even eigenmodes.

%
%

\end{document}